\documentclass[english]{article}
\usepackage[T1]{fontenc}
\usepackage[latin9]{inputenc}
\usepackage{geometry}
\usepackage{color}
\usepackage{babel}
\usepackage{amsbsy}
\usepackage{amstext}
\usepackage{amssymb}
\usepackage{cancel}
\usepackage{graphicx}
\usepackage{esint}
\usepackage[unicode=true,pdfusetitle,
 bookmarks=true,bookmarksnumbered=false,bookmarksopen=false,
 breaklinks=false,pdfborder={0 0 1},backref=false,colorlinks=false]
 {hyperref}

\makeatletter

\providecommand{\tabularnewline}{\\}

\newcommand{\lyxaddress}[1]{
	\par {\raggedright #1
	\vspace{1.4em}
	\noindent\par}
}

\makeatother

\begin{document}
\title{Analysis of grid instabilities in particle-in-cell codes based on
a meshfree approach}
\author{John M.~Finn$^{*}$ and Evstati G.~Evstatiev$^{**}$}
\maketitle

\lyxaddress{$^{*}$ Tibbar Plasma Technologies, 274 DP Rd., Los Alamos, NM 87544}

\lyxaddress{$^{**}$ Sandia National Laboratories, Albuquerque, NM 87185}
\begin{abstract}
In this paper, a new method of analyzing grid, or aliasing, instabilities
is described, in particle-in-cell (PIC) codes for plasma kinetic theory.
This method starts with a meshfree approach with $N_{p}$ macroparticles of width $w$
and leads to a simplification of the analysis, facilitating the analysis
for $N_{p}<\infty$. A method of linearizing these equations is described,
about a uniform density non-drifting equilibrium prescribed by macroparticles
on a uniform particle \emph{lattice} of dimension $N_{p}$. Perturbations
are induced by perturbing particle positions about this lattice. The
resulting linear equations are used to analyze meshfree stability
and dispersion relations. In the final step, the linearized equations
are discretized on a grid of $N_{g}$ points with $N_{g}<N_{p}$,
in both momentum conserving (MCP) and energy conserving (ECP) discretizations.
For a cold stationary plasma, this discretized linearization leads
to equations of motion with a $N_{p}\times N_{p}$ \emph{dynamical
matrix.} The ECP form is symmetric and positive definite for the uniform
density plasma, and immediately shows  stability. The MCP form of
the dynamical matrix is neither symmetric nor positive definite, preventing
immediate conclusions on stability. For the MCP and ECP discretizations
for a cold plasma and an integer number of lattice particles per cell\textbf{
$N_{ppc}=N_{p}/N_{g}\geq1$}, the resulting matrix elements vary relative
to the meshfree value, a result of aliasing. This deviation varies
periodically with the displacement of the particle lattice relative
to the grid. The block-circulant nature of the $N_{p}\times N_{p}$
ECP and MCP matrices, with blocks of size $N_{ppc}$, is potentially
useful in the analysis, but a simple transformation leads to a \emph{dimensional
reduction} to a $N_{g}\times N_{g}$ matrix which has the same nontrivial
eigenvalues and is \emph{circulant}, aiding further the analysis.
In the MCP discretization, eigenvalues occur in complex conjugate
pairs for general placement of the particle lattice relative to the
grid. These conjugate pairs indicate nonnegative growth rates which
vary periodically with the lattice displacement. The nature of the
aliasing is studied and shown to be related to the trapezoidal rule
error over the grid. The scaling of the linear growth rates with respect
to $N_{g}$ and especially with respect to $N_{ppc}$ is studied.
These analytical results are compared with PIC simulations and found
to be in excellent agreement. This variation of the matrix elements
with the lattice displacement for a cold non-drifting plasma is closely
related to the periodic variation in time for the matrix elements
for both MCP and ECP discretizations of a cold drifting \emph{beam}.
Issues related to this relationship, specifically grid--induced instabilities
for both the MCP and ECP approaches, are discussed briefly.

\newpage{}

\tableofcontents{}

\newpage{}
\end{abstract}

\section{Introduction\label{sec:Introduction}}

This paper focuses on particle-in-cell (PIC) computational methods
for plasma kinetic theory, using macroparticles. We concentrate on
an electrostatic one-dimensional formulation with immobile ions, periodic boundary
conditions, and net neutrality, i.e.~overall cancellation of the
electron and ion charges. We begin with a meshfree (MF) formulation.
One point of this paper is to establish methods of dealing with grid
instabilities in particle--in--cell (PIC) methods, both of the momentum
conserving (MCP) and the energy conserving (ECP) discretizations.
The approach of starting with a MF formulation and adding the grid
in the final step aids in the analysis and interpretation of the grid
effects.

Another point of this paper is to study grid instabilities that can
occur for a cold, non-drifting plasma, and to present the properties
of these instabilities, especially the scaling with respect to the
number of cells $N_{g}$ and the number of particles per cell $N_{ppc}=N_{p}/N_{g}$.

Our MF analysis for a cold plasma starts with specified particle positions,
with the electron density computed by kernel density estimation\cite{Bishop},
a well--known method of estimating a probability density based on
the location of samples. The translationally invariant kernel $K(x,y)=K(x-y)$,
called the \emph{particle shape} in PIC texts, represents the distribution
of charge density inside each of the $M$ macroparticles, hereafter
\emph{particles}. From the electron density computed in this manner,
the electric field can be computed in a similar manner. The force
on a particle is found by integrating the electric field over the
same kernel (macroparticle charge density) as for finding the density;
this is dictated by the basic source-target symmetry of electrodynamics,
and is a natural choice.

Based on the expression for the force, a method of linearization of
the equations of motion is described, with an equilibrium specified
by having particles on a uniformly spaced particle lattice $\{\xi_{\alpha0}\}_{\alpha=0}^{N_{p}-1}$
and equilibrium velocities $u_{\alpha0}$. For a cold stationary plasma these
are $u_{\alpha0}=0$. These produce an equilibrium with uniform electron
density, zero velocity (and temperature), and zero electric field.
(This particle lattice is not to be confused with the computational
grid $\{x_{i}\}$, to be introduced.) Initial perturbations are induced
by specifying perturbed particle positions $\delta\xi_{\alpha}$ relative
to $\xi_{\alpha0}$, and perturbed velocities $\delta u_{\alpha}$.
The resulting expressions lead to equations of motion for each of
the particles, and can be used for deriving dispersion relations and
for analyzing stability.

Dispersion relations for a cold plasma and a cold drifting beam are
derived by the above MF approach, and the limit $N_{p}\to\infty$
leads to the familiar Langmuir dispersion relation for these special
cases, with modifications due to the kernel shape and width. The $N_{p}<\infty$
expressions lead to first order correction (in $q=1/N_{p}$) to these
dispersion relations due to the finite number of particles and their
shape.

The final step is to formulate the equations discretized to a grid
$\{x_{i}\}_{i=0}^{N_{g}-1}$ of $N_{g}$ points, with grid spacing
$\Delta=1/N_{g}$, in both the momentum conserving (MCP) and energy
conserving (ECP) cases. It is observed  that for a cold stationary
plasma, the linearization for the ECP case leads to equations of motion
with a symmetric, positive definite (SPD) dynamical matrix, proving
stability. In the MCP discretization a similar linearized set of equations
of motion is derived, but the resulting matrix is neither symmetric
not positive definite, so stability cannot immediately be inferred. 

We specialize to cases for which the number of particles per cell
$N_{ppc}=N_{p}/N_{g}$ is a positive integer, and unless specified
otherwise consider a cold non-drifting plasma. With the grid fixed,
we introduce a uniform lattice displacement $\xi_{\alpha0}\to\xi_{\alpha0}+\epsilon$
and find a dependence of the resulting matrix elements on the displacement
$\epsilon$.  This variation for $N_{g}<\infty$ is traced to the
force on a particle being expressed as a trapezoidal rule sum over
the grid, an approximation of the integral used for a pure MF treatment,
and the trapezoidal rule error in the dynamical matrix elements is
seen to oscillate as $\epsilon$ is varied, with period $\Delta$.
Equivalently, the Fourier series of the force is broadband for localized
or nonsmooth kernels, and therefore the discrete Fourier transform
(DFT) has \emph{aliasing} which oscillates with $\epsilon$, with
period $\Delta$. For the MCP case, this trapezoidal or aliasing error
leads to complex eigenvalues (with growth rates depending on $\epsilon$)
indicating instability. In the ECP discretization, on the other hand,
the SPD nature of the matrix holds in spite of the aliasing errors,
guaranteeing stability. We find that the nontrivial eigenvalues of
the dynamical matrix for ECP are degenerate (and of course real),
as are the eigenvalues of the symmetric part of $M$ for MCP. This
is related to the fact that the matrix $M$ for the MCP case has complex
conjugate eigenvalues and therefore has positive growth rates $\gamma$
for any $N_{g}$. These growth rates are periodic of period $q=1/N_{p}$.
These analytical results are compared, and show agreement, with PIC
simulations.

This variation of the matrix elements with the lattice displacement
for a cold plasma is closely related to the variation in time for
the matrix elements for both MCP and ECP discretizations for a cold
\emph{beam}. Issues related to this relationship, specifically grid--induced
instabilities for both the ECP and MCP approaches, are discussed briefly,
and will be dealt with in detail in a forthcoming publication.

In Sec.~\ref{sec:Meshfree-formulation} we introduce a specific form
of meshfree equations, based on a kernel $K(x-y)$ to find the electron
density and a related function $G(x-y)$ used to find the electric
field, giving a Vlasov--Gauss formulation. We discuss the charge,
momentum and energy conservation laws associated with this formulation,
and discuss the equivalence of this Vlasov--Gauss formulation with
macroparticles with the Vlasov--Poisson and Vlasov--Amp\`ere formulations.

In Sec.~\ref{sec:Linearization-of-Meshfree}\label{Fig:dkernel-gkernel-1}
we describe a method of linearizing the MF equations about an equilibrium,
specialized to the constant electron density case. This equilibrium
has a uniformly spaced particle lattice $\{\xi_{\alpha0}\}$, providing
a uniform density equilibrium which, with uniform ion density, leads
to zero electric field. The field is perturbed by introducing initial
perturbations of the particles from the lattice positions. Results
for a cold plasma and a cold beam, specifically in the form of dispersion
relations, are derived and discussed.

In Sec.~\ref{sec:Discrete-equations-of} we describe discretization
of the MF equations for both the MCP and ECP formulations. In the
latter, the linearized equations for a cold plasma are written in
terms of a symmetric, positive definite $N_{p}\times N_{p}$ \emph{dynamical
matrix} $M$, immediately implying stability. For the MCP case the
corresponding dynamical matrix is neither symmetric nor positive definite,
and no immediate stability conclusions are possible.

In Sec.~\ref{sec:Properties-of-the} we discuss discretization of
the $N_{p}\times N_{p}$ matrices $M$ for both the ECP and MCP cases,
and the error relative to the MF version. This error can be characterized
as trapezoidal rule error or, equivalently, the aliasing error. We
also discuss the block-circulant nature of these matrices if $N_{ppc}=N_{p}/N_{g}>1$.
More importantly, we construct a smaller, $N_{g}\times N_{g}$, associated
dynamical matrix $\hat{M}$. We show that $\hat{M}$ is circulant,
allowing discrete Fourier transform (DFT) analysis. The eigenvalues
of the $N_{g}\times N_{g}$ matrix $\hat{M}$ are also eigenvalues
of the $N_{p}\times N_{p}$ dynamical matrix $M$, which in addition
has $N_{p}-N_{g}$ trivial zero eigenvalues; removing these trivial
eigenvalues is a major point of using the dimensionally reduced matrix
$\hat{M}$. In this section we discuss briefly the possibility that
the dynamical matrix (but not the reduced dimension dynamical matrix)
is non-normal, allowing for transient growth.

In Sec.~\ref{sec:Shifted-lattice} we discuss the application of
a rigid shift of the lattice points, $\xi_{\alpha0}\to\xi_{\alpha0}+\epsilon$,
showing the relation between the degeneracy of the real eigenvalues
of the symmetric part of $M$ in its MCP form and the growth rate
(imaginary part of eigenvalues) obtained from the full MCP form of
the matrix. 

In Sec.~\ref{sec:Cold-beam} we describe the form of the matrices
for a cold drifting beam, pointing out a connection between the static
displacement of the lattice and a cold beam. We briefly discuss how
these cold beam results can by used to study the stability of a warm
plasma. Results related to these cases will be presented in more detail
in forthcoming publications.

In Sec.~\ref{sec:Summary-and-Discussion} we summarize and discuss
our results.

In Appendix \ref{app:Trapezoidal} we show details related to trapezoidal
rule error, and show that it is equivalent to aliasing error.\textbf{
}In Appendix \ref{app:DimReduction} we discuss some details regarding
the reduction in the size of the dynamical matrix to $N_{g}\times N_{g}$;
this includes the fact that this matrix is circulant. This can be
important if $N_{p}\gg N_{g}$. We also briefly discuss the equivalence
of this dynamical reduction and the derivation of fluid moments discretized
to the grid.

\section{Meshfree formulation}
\label{sec:Meshfree-formulation}

In this paper we focus on 1D kinetic theory for high frequency phenomena,
i.e.~with immobile ions with the ion number density $n_{i0}$ uniform;
the electron number density is written as $n_{e}$ with mass density
$\rho_{e}=n_{e}m_{e}$ and charge density $-n_{e}e$. We consider
periodic boundary conditions of period $L$, and a net neutral plasma,
with $\int_{0}^{L}n_{e}(x)dx=n_{i0}L$. The constant density equilibrium
has $n_{e0}=n_{i0}$ as well as $E(x)=0$. Time is normalized by $t\to\omega_{pe}t$,
with $\omega_{pe}^{2}=n_{e0}e^{2}/m_{e}\epsilon_{0}$, space by $x\to x/L$
(i.e.~$L=1$), velocities to $L\omega_{pe}$, charge density to $n_{i0}e$,
electron mass density to $n_{i0}m_{e}$, and electric field to $n_{e0}eL/\epsilon_{0}$.
In a meshfree formulation with $N_{p}$ macroparticles located at
positions $\xi_{\alpha}$, $\alpha=0,\dots,N_{p}-1$, the dimensionless
electron charge density $\rho$ (without the minus sign) and the electron
mass density are equal and satisfy $\int_{0}^{1}\rho(x)dx=\rho_{i0}=1$.
In these units, where the net charge density and mass are both unity,
the macroparticle charge is $q=1/N_{p}$ and the macroparticle mass
$m$ equals $q$.

As useful as the meshfree equations for this analysis and for the
discretization to a grid in Sec.~\ref{sec:Discrete-equations-of},
there are serious impediments to using a meshfree code. These are
traced to the fact that the computation time for such a code scales
as $N_{p}^{2}$. These disadvantages might be ameliorated by picking
a low discrepancy distribution\cite{lemieux2009monte} of particle
positions (for a cold plasma) or particle positions and velocities
(for a warm plasma.) For such a distribution the amplitude of the
particle noise can scale as $1/N_{p}$ rather than $1/N_{p}^{1/2}$
for independent identically distributed (i.i.d.) sampling. This allows
for the use of fewer particles. Such a low-discrepancy sampling is
of course possible for the initial conditions. Moreover, for runs
in the linear regime, where particles do not move much, this may be
adequate. However, for nonlinear behavior the particles move sufficiently
to lose the low discrepancy property, and serious modifications would
be required to continue.

\subsection{Meshfree equations of motion}
\label{subsec:Meshfree-equations-of}

We write the unsigned electron density in terms of the macroparticle
positions as
\begin{equation}
\rho(x,t)=q\sum_{\beta=0}^{N_{p}-1}K(x-\xi_{\beta}(t))\label{eq:KDE}
\end{equation}
with 
\begin{equation}
q=1/N_{p}\text{ and }\int K(x)dx=1.\label{eq:q=00003D,intK=00003D1}
\end{equation}
In this normalization the macroparticle unsigned charge and the electron
mass are both $q$, so that Eq.~(\ref{eq:KDE}) also represents the
electron mass density.

At this stage we will assume that the plasma is cold and the initial
macroparticle velocities are zero. The kernel $K$ represents the
charge density of the macroparticle centered at $\xi_{\alpha}$ and
is called the \emph{particle shape} \cite{BLBook,HockneyEastwood}.
This kernel has the translationally invariant form $K(x,y)=K(x-y)$,
consistent with the uniformity of $\rho_{i0}$ and the periodic boundary
conditions. Also, the kernel $K$ is extended to be periodic of period
unity. For convenience we assume $K(x)\geq0$ and take $K$ to have
compact support with full width $w$, with periodicity in (dimensionless)
$x$ of period $1$. Equations (\ref{eq:KDE}) and (\ref{eq:q=00003D,intK=00003D1})
lead to
\begin{equation}
\int\rho(x,t)dx=N_{p}q=1,\label{eq:total-e-charge}
\end{equation}
as required\footnote{The process described in Eq.~(\ref{eq:KDE}) is called \emph{kernel
density estimation} in the statistics and Machine Learning communities\cite{Bishop}.
From this point of view, the $\xi_{\alpha}$ are samples from a given
density and $\rho$ is its estimate.}. Gauss's law in these normalized variables is
\begin{equation}
E'(x,t)=1-\rho(x,t),\label{eq:Gauss's-law}
\end{equation}
where we take the sign of the electron charge to be negative explicitly.
We also take the potential difference over a period $\Delta\phi=-\int_{0}^{1}E(x)dx$
to be zero. We express the electric field in terms of the field $qG(x-\xi_{\beta},t)$
due to a single macroparticle of width $w$ at $\xi_{\beta}$ with
its share of neutralizing ion charge,
\begin{equation}
E(x,t)=q\sum_{\beta}G(x-\xi_{\beta}(t)).\label{eq:E(x)-rep-G(x)}
\end{equation}
We write $G(x)$ in terms of $G_{0}(x)$, the Green's function for
Gauss's law including the ion charge $\rho_{i}=\rho_{i0}=1$, requiring
\begin{equation}
G_{0}'(x)=1-\delta(x).\label{eq:ODE-for-GF-G0}
\end{equation}
Using the zero electrostatic potential difference condition $\int G_{0}(x)dx=0$,
this implies 
\begin{equation}
G_{0}(x)=x-\sigma(x)/2,\label{eq:G0-Greens-function}
\end{equation}
where $\sigma(x)$ is the sign of $x$. Neutrality leads to the conditions
$G_{0}(1)=G_{0}(0)$. Then, defining $G$ by smoothing over the finite
macroparticle charge density,
\begin{equation}
G(x)=\int G_{0}(z)K(x-z)dz=\int G_{0}(x-y)K(y)dy,\label{eq:G=00003DG0,smoothed}
\end{equation}
we find, using Eq.~(\ref{eq:q=00003D,intK=00003D1}), that $G$ satisfies
\begin{equation}
G'(x)=1-K(x)\label{eq:G'=00003D1-K}
\end{equation}
with 
\begin{equation}
G(1)=G(0);\,\,\,\int_{0}^{1}G(x)dx=0.\label{eq:Zero-pot-diff-G(x)}
\end{equation}

\begin{figure}
  \centering
\includegraphics[width=0.4\textwidth]{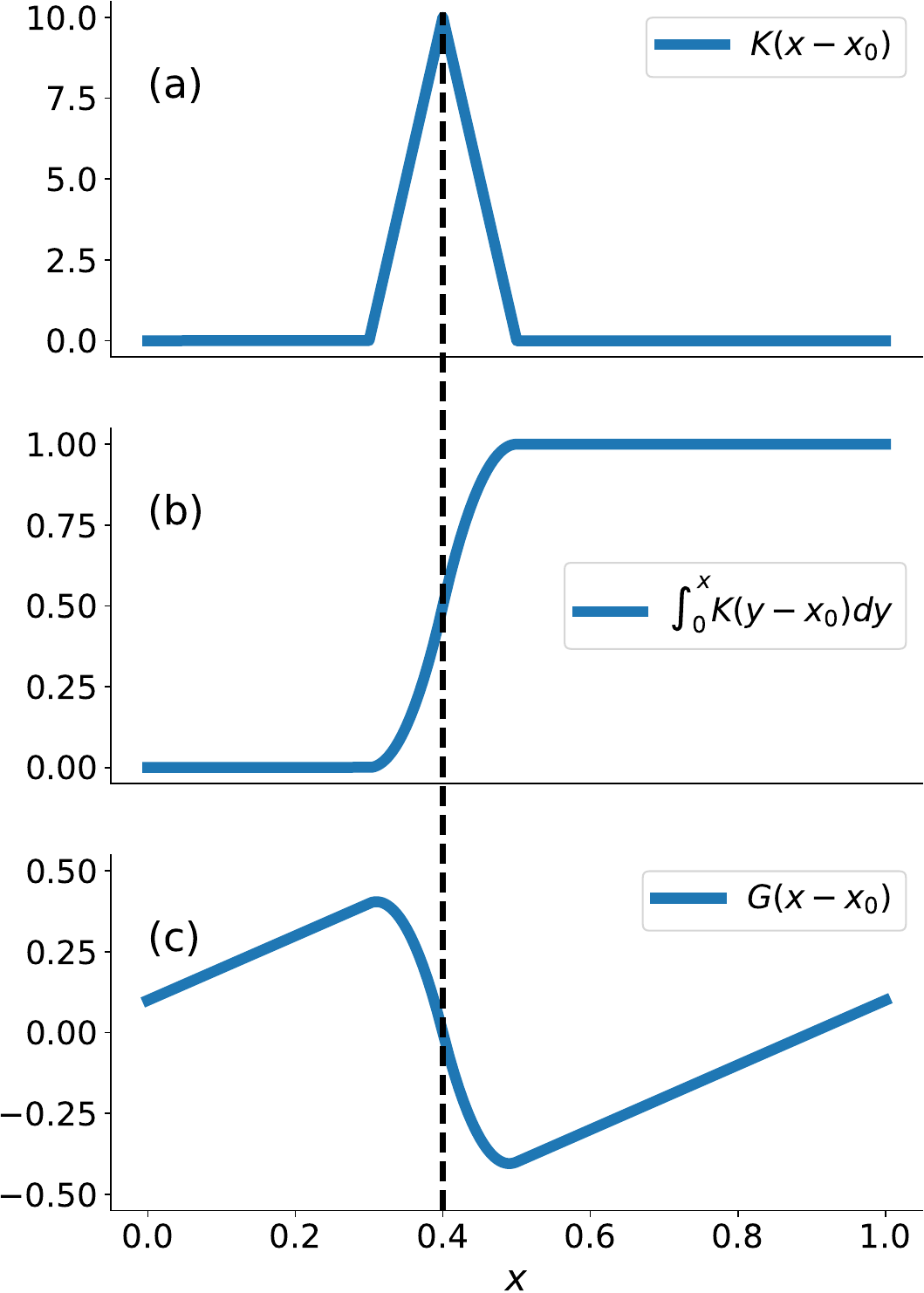}    
\caption{Density estimation kernel (a) $K(x-x_{0})$ for $x_{0}=0.4$ and full
width $w=0.2$, (b) $\int_{x_{0}}^{x}K(y)dy$ and (c) electric field
kernel $G(x-x_{0})$ . For the full width satisfying $w<1$ and compact
support, the periodic form of $K$ has $K(x)\to\sum_{n=-\infty}^{\infty}K(x-n)$
and similarly for $G$.}
\label{Fig:dkernel-gkernel}
\end{figure}

Defining
\begin{equation}
E(x,t)=\int G_{0}(x-y)\rho(y,t)dy,\label{eq:E(x)-from-G0}
\end{equation}
we conclude that Gauss's law for electric field given by Eq.~(\ref{eq:E(x)-rep-G(x)})
is indeed satisfied,
\begin{equation}
E'(x,t)=q\sum_{\beta}\left(1-K(x-\xi_{\beta}(x,t))\right)=1-\rho(x,t).\label{eq:Gauss-indeed-sat}
\end{equation}
Examples of a kernel $K(x)$ and the field kernel $G(x)$ (both with
full width $w=0.2$) are shown in Fig.~\ref{Fig:dkernel-gkernel}.
The zero potential difference condition\cite{OurJCP} $\Delta\phi\doteq-\int E(x,t)dx=0$
is satisfied because of condition (\ref{eq:Zero-pot-diff-G(x)}).
(In the remainder of this paper, integrals over $[0,1]$ will have
limits implicit.) We often assume $K$ to be symmetric (even), $K(-x)=K(x)$,
and in this case we can define 
\begin{equation}
G(x)=x-\int_{0}^{x}K(y)dy\label{eq:G-for-K-even}
\end{equation}
with $G$ odd. However, as we will see, there is no need to assume
this symmetry condition on the `bare' kernel $K$.

The force \footnote{This force at $\xi_{\alpha}$ depends on all the $\xi_{\beta}$, as
shown in Eq.~(\ref{eq:Force-at-beta}), so the notation $F_{\alpha}$
avoids the confusion of writing $F(\xi_{\alpha})$.} on another macroparticle at $\xi_{\alpha}$ is found by integrating
the negative of the electric field over the macroparticle charge density
using the same kernel $K$,
\begin{equation}
F_{\alpha}=-q\int E(y)K(y-\xi_{\alpha})dy\label{eq:Force-at-beta-0}
\end{equation}
\begin{equation}
=-q^{2}\sum_{\beta}\int G(y-\xi_{\beta})K(y-\xi_{\alpha})dy.\label{eq:Force-at-beta}
\end{equation}
(Recall, according to the normalizations above -- unsigned macroparticle
charge and macroparticle mass $m$ are both equal to $q$.) The force
on a macroparticle at $\xi_{\alpha}$ is
\begin{equation}
F_{\alpha}=-q^{2}\sum_{\beta}G_{2}(\xi_{\alpha}-\xi_{\beta}).\label{eq:Force-at-beta-2}
\end{equation}
Here we define
\begin{equation}
G_{2}(\xi_{\alpha}-\xi_{\beta})=\int G(y-\xi_{\beta})K(y-\xi_{\alpha})dy=\int G(z+\xi_{\alpha}-\xi_{\beta})K(z)dz.\label{eq:G2-definition}
\end{equation}
From the last form we write $\int_{0}^{1}G_{2}(x)dx=\int dzK(z)\int dxG(z+x)$
and therefore
\begin{equation}
\int_{0}^{1}G_{2}(x)dx=0\label{eq:G2-loop-too}
\end{equation}
and conclude further
\[
G_{2}'(\xi_{\alpha}-\xi_{\beta})=\int G'(z+\xi_{\alpha}-\xi_{\beta})K(z)dz
\]
\[
=\int\left(1-K(z+\xi_{\alpha}-\xi_{\beta})\right)K(z)dz,
\]
\begin{equation}
=1-K_{2}(\xi_{\alpha}-\xi_{\beta})\label{eq:G2'=00003D1-K2}
\end{equation}
where $\int K(x)dx=1$ has been used. The form in Eq.~(\ref{eq:G2'=00003D1-K2})
is the analog of Eq.~(\ref{eq:G'=00003D1-K}), where 
\begin{equation}
K_{2}(\xi_{\alpha}-\xi_{\beta})=\int K(z+\xi_{\alpha}-\xi_{\beta})K(z)dz=\int K(v-\xi_{\beta})K(v-\xi_{\alpha})dv,\label{eq:K2-2-forms}
\end{equation}
the correlation of $K$ with itself, $K_{2}=K\star K$. In this form
the properties of translation invariance, symmetry and positive definiteness
are evident. In discrete form this is the kernel matrix $K_{2\alpha\beta}=\Delta\sum_{i}K(x_{i}-\xi_{\alpha})K(x_{i}-\xi_{\beta})$,
of the form $\Delta\sum_{i}K_{\alpha i}^{T}K_{i\beta}$, a Gram matrix,
which is symmetric and positive definite. Also, $K_{2\alpha\beta}$
is circulant, the discrete analog of translational invariance, if
$K_{\alpha\beta}$ is. We have noted that the symmetry of $K_{2}$
exists because of the basic symmetry between source and target charges
in electrodynamics, i.e.~the kernels used in Eqs.~(\ref{eq:KDE})
and (\ref{eq:Force-at-beta-0}) are necessarily identical\footnote{The equality of the two kernels, for the source of electric field and
for the force on a target particle, is physically motivated and automatically
yields a SPD kernel $K_{2}$. (This process is not appropriately defined
as an interpolation of the electric field.) The derivation of $K_{2}$
from $K$ is similar to the way that a SPD (reproducing) kernel is
defined in terms of feature space vectors in statistics and Machine
Learning\cite{Bishop}.}. The properties in Eqs.~(\ref{eq:G2-loop-too}) and (\ref{eq:G2'=00003D1-K2})
show that $G_{2}(x)$ is odd, even if $G(x)$ is not, and this leads
to the property $G_{2}(0)=0$, which implies that the self--force
is zero, so the term in Eq.~(\ref{eq:Force-at-beta-2}) with $\beta=\alpha$
is zero.

Summarizing, we conclude that $K_{2}$ is symmetric and positive definite,
and that $G_{2}$ is antisymmetric (odd). We do not need to require
the kernel $K$ in Eqs.~(\ref{eq:KDE}) and (\ref{eq:Force-at-beta-0})
to be either symmetric or positive definite. In the language of Statistics
(Machine Learning) \cite{Bishop}, $K$ is a feature in the feature
space, and its correlation $K_{2}$ is the positive definite kernel.

\subsection{Exact equations for $N_{p}$ macroparticles; conservation laws}
\label{subsec:Exact-equation-for}

From the relations above, we can write the equations of motion of
the $N_{p}$ macroparticles, 
\begin{equation}
\frac{d\xi_{\alpha}}{dt}=u_{\alpha},\label{eq:EM-1}
\end{equation}
\begin{equation}
m\frac{du_{\alpha}}{dt}=F_{\alpha}=-q^{2}\sum_{\beta;\beta\neq\alpha}G_{2}(\xi_{\alpha}-\xi_{\beta}),\label{eq:EM-2}
\end{equation}
where, as we have discussed in the last section, $F_{\alpha}$ is
the force on the particle at $\xi_{\alpha}$. We write
\begin{equation}
H=m\sum_{\alpha}\frac{u_{\alpha}^{2}}{2}+\sum_{\alpha,\beta;\alpha<\beta}V(\xi_{\alpha}-\xi_{\beta}),\label{eq:Hamiltonian}
\end{equation}
where $q$ here is the macroparticle mass. The condition $\alpha<\beta$
holds over the circle in a cyclic manner; for example, for $N_{p}=3$
we have $V(\xi_{1}-\xi_{2})+V(\xi_{2}-\xi_{3})+V(\xi_{3}-\xi_{1})$.
 We conclude that $H$ is the Hamiltonian for the system in Eqs.~(\ref{eq:EM-1})
and (\ref{eq:EM-2}), in canonical variables $(\xi_{\alpha},u_{\alpha})$,
if $V$ is the potential energy for the force in Eq.~(\ref{eq:EM-2}),
i.e.
\[
V'(\xi_{\alpha}-\xi_{\beta})=q^{2}G_{2}(\xi_{\alpha}-\xi_{\beta}).
\]

Let us turn to the conservation laws of the equations of motion in
Eqs.~(\ref{eq:EM-1}) and (\ref{eq:EM-2}). First of all, from Eq.~(\ref{eq:KDE})
the continuity equation holds exactly. Indeed, this equation implies
\begin{equation}
\partial_{t}\rho(x,t)=-q\sum_{\beta}K'(x-\xi_{\beta}(t))\dot{\xi}_{\beta}(t)\label{eq:Continuity1}
\end{equation}
\begin{equation}
=-\partial_{x}\left[q\sum_{\beta}K(x-\xi_{\beta}(t))u_{\beta}(t)\right]=-\partial_{x}\left[\rho(x,t)u(x,t)\right],\label{eq:Continuity2}
\end{equation}
Where 
\begin{equation}
\rho(x,t)u(x,t)=q\sum_{\beta}K(x-\xi_{\beta}(t))u_{\beta}(t).\label{eq:Def-u(x,t)}
\end{equation}
Writing the conditional probability density $f(v|x,t)$ as the ratio
of the joint density $f(x,v,t)$ to the marginal density $\rho(x,t)$,
i.e.~$f(v|x,t)=f(x,v,t)/\rho(x,t)$, we can write the joint phase
space density in the usual manner\cite{BLBook}
\begin{equation}
f(x,v,t)=q\sum_{\beta}K(x-\xi_{\beta}(t))\delta(v-u_{\beta}(t)),\label{eq:Joint-and-marginal}
\end{equation}
and the density is the marginal distribution
\begin{equation}
\rho(x,t)=\int f(x,v,t)dv,\label{eq:Marginal}
\end{equation}
consistent with Eq.~(\ref{eq:KDE}). Using Eqs.~(\ref{eq:Joint-and-marginal})
and (\ref{eq:Marginal}) we obtain
\begin{equation}
f(v|x,t)=\frac{q\sum_{\beta}K(x-\xi_{\beta}(t))\delta(v-u_{\beta}(t))}{\rho(x,t)},\label{eq:Conditional}
\end{equation}
showing that the conditional expectation
\begin{equation}
\int f(v|x,t)vdv\label{eq:Cond-expect}
\end{equation}
equals the velocity $u(x,t)$ in Eq.~(\ref{eq:Def-u(x,t)})\footnote{In the Statistics community, this estimate based on the conditional
expectation is called the Nadaraya--Watson estimate\cite{Bishop}.}. 

We conclude that the continuity equation, Eqs.~(\ref{eq:Continuity1})
and (\ref{eq:Continuity2}) with $u(x,t)$ given by Eq.~(\ref{eq:Def-u(x,t)}),
is satisfied exactly. This immediately implies that the total electron
charge
\begin{equation}
Q=\int\rho(x,t)dx\label{eq:TotalCharge}
\end{equation}
and the total electron mass are conserved exactly, and the total charge
remains zero.

Conservation of momentum is demonstrated by the symmetry $\xi_{\beta}\to\xi_{\beta}+\epsilon\text{ mod}(1)$
of the Hamiltonian in Eq.~(\ref{eq:Hamiltonian}) (or its associated
Lagrangian). Also, we can conclude directly from Eq.~(\ref{eq:EM-2})
that the momentum $P$, equal to $\int\rho udx=m\sum_{\alpha}u_{\alpha}$,
satisfies
\begin{equation}
\dot{P}=m\sum_{\alpha}\dot{u}_{\alpha}=-q^{2}\sum_{\alpha,\beta}G_{2}(\xi_{\alpha}-\xi_{\beta})=0.\label{eq:MomentumCons}
\end{equation}
For the last point, we noticed at the end of Sec.~\ref{sec:Meshfree-formulation}
that $G_{2}$ is odd, allowing us to conclude that momentum is conserved.

Relaxing the $\Delta\phi=0$ condition for a moment, we can write,
without invoking $\Delta\phi=0$,
\[
\dot{P}=\sum_{\alpha}F_{\alpha}=-q\sum_{\alpha}\int dyE(y)K(y-\xi_{\alpha}).
\]
Then using Eq.~(\ref{eq:KDE}) we find $\dot{P}=\int\phi'(y)\rho(y)dy$.
Using $\phi''(y)=\rho(y)-1$, this leads to
\begin{equation}
\dot{P}=\Delta\phi+\left[\frac{\phi'(x)^{2}}{2}\right]_{0}^{1}=\Delta\phi;\label{eq:P-dot-w/o=00005CDeltaphi=00003D0}
\end{equation}
here, the condition $E(1)=E(0)$ has been used. The condition $\Delta\phi=0$
shows agreement with Eq.~(\ref{eq:MomentumCons}). The condition
$\dot{P}=\Delta\phi$ in Eq.~(\ref{eq:P-dot-w/o=00005CDeltaphi=00003D0})
will be used later in this section. 

We note that the energy $H$ in Eq.~(\ref{eq:Hamiltonian}) is conserved
in continuous time because of the Hamiltonian nature of the equations,
and their lack of explicit time dependence. Let us again consider
finite time--stepping of Eqs.~(\ref{eq:EM-1}),(\ref{eq:EM-2}).
Because of the Hamiltonian nature of these equations (in canonical
variables), there are several symplectic integrators that can be used.
Among these are implicit midpoint (IM)\cite{Hairer}, in the linear
context called Crank--Nicolson, and symmetrized symplectic Euler\cite{Hairer,MyIssues}.
The latter scheme, with a separable Hamiltonian as in Eq.~(\ref{eq:Hamiltonian}),
reduces to symmetrized leapfrog (Verlet)\cite{Hairer,Josh-Leland-2nd-order,Leland,MyIssues,OurJCP,Richardson-Adaptive}
These integrators are second order accurate. The typical behavior
of such second order accurate symplectic integrators is that for a
small enough time step $h$, the energy may not be exactly conserved,
but there is a modified energy function $H'=H+O(h^{2})$ that \emph{is}
exactly conserved. The actual energy $H$ is seen to oscillate within
bounds with amplitude $O(h^{2})$. This avoids the main problem of
non--symplectic schemes, namely allowing unbounded variation of energy
with time.

These conservation properties derive from the fact that the equations
of motion, Eqs.~(\ref{eq:EM-1}) and (\ref{eq:EM-2}), described
by the Hamiltonian, Eq.~(\ref{eq:Hamiltonian}), are the exact equations
of motion for the macroparticles as rigid clouds of charge that do
not interact mechanically. The approximation that the original system
is described in terms of macroparticles of width $w$ preserves the
Hamiltonian nature of the full system of actual charges.

\subsection{Vlasov--Gauss, Vlasov--Poisson, and Vlasov--Amp\`ere formulations}
\label{subsec:Vlasov-Gauss-=00003D-Vlasov-Gauss}

The meshfree approach presented in this section in terms of the electric
field can be described as a Vlasov--Gauss (VG) formulation. The Vlasov--Poisson
(VP) formulation is identical because the relation between the potential
$\phi$ and the electric field is in terms of actual spatial derivatives
rather than a discretization. That is, we can write the potential,
with $E=-\phi'$, as
\begin{equation}
\phi(x,t)=q\sum_{\beta}\Phi(x-\xi_{\beta}(t)),\label{eq:phi-particle-rep}
\end{equation}
where $\Phi(x-\xi_{\beta})$ is the potential due to the particle
at $\xi_{\beta}$ with its share of neutralizing ion charge. That
is, we assume
\begin{equation}
G(x)=-\Phi'(x)\label{eq:G=00003D-phi'}
\end{equation}
with zero potential difference condition $\Phi(1)=\Phi(0)$. If we
use Eq.~(\ref{eq:E(x)-rep-G(x)}) we indeed conclude $E(x,t)=-\phi'(x,t)$
and $\phi(1,t)=\phi(0,t)$, i.e.~$\Delta\phi=-\int E(x,t)dx=0$.
There is also a gauge invariance $\phi\to\phi+C$, a null space dimension
when dealing with the potential $\phi$, resolved by the condition
$\Phi(1)=\Phi(0)=0$.

Let us consider the issue of equivalence between the Vlasov--Amp\`ere
(VA) formulation and the VG and VP formulations, as long as the formulation
is in terms of time derivatives rather than finite time--stepping.
In the VA approach in 1D and continuous time with $N_{p}$ macroparticles,
we integrate
\begin{equation}
\partial_{t}E(x,t)=-j(x,t),\label{eq:VA-equation}
\end{equation}
where
\begin{equation}
j(x,t)=-\rho(x,t)u(x,t)=-q\sum_{\alpha=1}^{N_{p}}K(x-\xi_{\alpha}(t))u_{\alpha}(t)\label{eq:VA-Current}
\end{equation}
is the current density. The fact that neither the spatial nor temporal
derivatives are discretized means that VA is equivalent to VG and
therefore to VP. Indeed, Eq.~(\ref{eq:VA-equation}) implies
\begin{equation}
\partial_{xt}E(x,t)=-\partial_{x}j(x,t)=\partial_{x}\left(\rho(x,t)u(x,t)\right)=-\partial_{t}\rho,\label{eq:d_x-Ampere}
\end{equation}
because, as discussed above, the continuity equation is satisfied
exactly. We find, as usual, that Gauss's law is satisfied for all
$t$,
\begin{equation}
\partial_{x}E(x,t)=1-\rho(x,t),\label{eq:Gauss'sLaw<-Ampere}
\end{equation}
as long it is satisfied at $t=0$. The neutralizing effect of the
stationary ions, i.e.~$\int(1-\rho(x,t))dx=0$, is consistent with
the boundary condition $E(1)=E(0)$. For the final boundary condition
$\Delta\phi=0$, we show that the potential difference satisfies $(d/dt)\Delta\phi(t)=-\int\partial_{t}E(x,t)dx=-\int\rho udx=-P$.
But we have also Eq.~(\ref{eq:P-dot-w/o=00005CDeltaphi=00003D0}),
i.e.~$(d/dt)P=\Delta\phi$. Putting these together, we find that
if $\Delta\phi=0$ and $P=0$ at $t=0$, the same condition holds
for all $t>0$. Moreover, the initial state with $\Delta\phi=P=0$
is stable.

We conclude that the VG, VP, and VA formulations are equivalent for
the MF equations with a finite number $N_{p}$ of macroparticles.
But recall that we have assumed at this stage that the time and spatial
derivatives are not discretized, except for keeping a finite number
$N_{p}$ of macroparticles of full width $w$.

\section{Linearization of Meshfree equations}
\label{sec:Linearization-of-Meshfree}

To linearize the equations of motion, using Eq.~(\ref{eq:Force-at-beta-2})
and perturbing Eqs.~(\ref{eq:EM-1}) and (\ref{eq:EM-2}), we introduce
a uniformly spaced lattice of equilibrium macroparticle positions
$\{\xi_{\alpha0}\}$ for $\alpha=0,\dots,N_{p}-1$, i.e.~$\xi_{\alpha0}=\alpha\Delta_{p}$,
again with $\Delta_{p}=q=1/N_{p}$. We will begin in Sec.~\ref{subsec:LinearizationForaCold}
by assuming that the equilibrium distribution has uniform density
and is cold and stationary, i.e.~$f_{0}(x,v)=\delta(v)$, or $u_{\alpha0}=0$.
We will briefly discuss plasmas with uniform density and a beam distribution
of velocities $f_{0}(x,v)=\delta(v-v_{0})$ in Sec.~\ref{subsec:Linearization-for-a}.
This uniform lattice of equilibrium points $\xi_{\alpha0}$ (with
$u_{\alpha0}=0$) leads, by Eq.~(\ref{eq:KDE}) to a uniform equilibrium
density $\rho_{0}(x)=1$. This condition is the \emph{partition of
unity} (POU) on the lattice $\{\xi_{\alpha0}\}$. This useful condition
is satisfied by the boxcar kernel of full width any integer multiple
of $q$. (We will return to the issue of POU on the grid later.) Also,
any number of convolutions of the boxcar with itself obeys this condition,
including the linear (tent) kernel, the quadratic spline kernel and
the cubic spline kernel.

To proceed with the linearization process, we perturb the positions
with $\xi_{\alpha}=\xi_{\alpha0}+\delta\xi_{\alpha}$, with perturbations
$\delta\xi_{\alpha}$ and obeying the condition $|\delta\xi_{\alpha}|<q/2$,
so that the particle positions $\xi_{\alpha}$ are initialized within
separate intervals of width $q$.\footnote{This sampling, for randomly selected samples $\delta\xi_{\alpha}$
from the uniform distribution on $[-q/2,q/2]$ is an example of the
\emph{jittered grid} low discrepancy sampling, of importance in Monte--Carlo
methods\cite{lemieux2009monte,Discrepancy}.}

\subsection{Linearization for a cold plasma}
\label{subsec:LinearizationForaCold}

For a cold plasma, perturbing Eqs.~(\ref{eq:EM-1}) and (\ref{eq:EM-2}),
we find $m\delta\ddot{\xi}_{\alpha}=\delta F_{\alpha}$, where $\delta F_{\alpha}$
is the perturbed force 
\begin{equation}
\delta\ddot{\xi}_{\alpha}=\delta F_{\alpha}/m=-q\sum_{\beta}G_{2}'(\xi_{\alpha0}-\xi_{\beta0})\left(\delta\xi_{\alpha}-\delta\xi_{\beta}\right).\label{eq:PertEOM-1}
\end{equation}
We assume that the full width of the macroparticles $w$ is an integer
multiple of $q$, the lattice spacing. (When we introduce a grid of
spacing $\Delta=1/N_{g}$ in the next section, we will assume $\Delta=N_{ppc}q$,
where $N_{ppc}=N_{p}/N_{g}$ is an integer. Then, the assumption $w=k\Delta$
for $k=1$ to $3$, leading to $w=kN_{ppc}q$. As discussed above,
$K$ satisfies the POU on the lattice $\{\xi_{\alpha0}\}$, i.e.~$q\sum_{\alpha}K(\xi_{\alpha0})=1$.
This implies that the same holds for $K_{2}$ \cite{OurJCP}, and
this implies $\sum_{\beta}G_{2}'(\xi_{\alpha0}-\xi_{\beta0})=\sum_{\beta}\left(1-K_{2}(\xi_{\alpha0}-\xi_{\beta0})\right)=0$
for all $\alpha$, so the terms proportional to $\delta\xi_{\alpha}$
on the right in Eq.~(\ref{eq:PertEOM-1}) sum to zero. This implies
\begin{equation}
\delta\ddot{\xi}_{\alpha}=q\sum_{\beta}G_{2}'(\xi_{\alpha0}-\xi_{\beta0})\delta\xi_{\beta}=-q\sum_{\beta}\left(K_{2}(\xi_{\alpha0}-\xi_{\beta0})-1\right)\delta\xi_{\beta}.\label{eq:PertEoM-2}
\end{equation}
The perturbed force in Eq.~(\ref{eq:PertEoM-2}) is zero for the
uniform displacement $\delta\boldsymbol{\xi}\propto\boldsymbol{c}_{p}=(1,1,\dots,1)_{N_{p}}$,
because $K_{2}$ satisfies the POU on the lattice $\{\xi_{\alpha0}\}$.
This neutral mode just rotates the particles uniformly on the circle
(on a period), leading to zero density perturbation and zero electric
field. The matrix multiplying $\delta\boldsymbol{\xi}$ in this equation
is $K_{2}-\boldsymbol{c}_{p}\boldsymbol{c}_{p}^{T}$, where here $K_{2}$
has  elements $K_{2\alpha\beta}=K_{2}(\xi_{\alpha0}-\xi_{\beta0})$
and $P=q\boldsymbol{c}_{p}\boldsymbol{c}_{p}^{T}$ is the projection
operator to the subspace spanned by the uniform vector $\boldsymbol{c}_{p}$.
The partition of unity shows $K_{2}\boldsymbol{c}_{p}=\boldsymbol{c}_{p}$,
and $K_{2}-\boldsymbol{c}_{p}\boldsymbol{c}_{p}^{T}$ projects out
components parallel to $\boldsymbol{c}_{p}$. In the subspace with
$\delta\boldsymbol{\xi}\perp\boldsymbol{c}_{p}$, we have $(K_{2}-\boldsymbol{c}_{p}\boldsymbol{c}_{p}^{T})\delta\boldsymbol{\xi}=K_{2}\delta\boldsymbol{\xi}$.
In the remainder we will assume that the perturbed vectors $\delta\boldsymbol{\xi}$
are in this subspace orthogonal to $\boldsymbol{c}_{p}$. For such
vectors we have $\delta\ddot{\xi}_{\alpha}=\delta F_{\alpha}/m$ or
\begin{equation}
\delta\ddot{\xi}_{\alpha}=-q\sum_{\beta}K_{2}(\xi_{\alpha0}-\xi_{\beta0})\delta\xi_{\beta}.\label{eq:PertEoM-3}
\end{equation}
This and the symmetric, positive definite nature\footnote{Mercer's theorem\cite{Paulsen} says that the kernel $K(x-y)$ is
positive definite if and only if $K_{2\alpha\beta}=K_{2}(\xi_{\alpha0}-\xi_{\beta0})$
is a positive definite matrix for all sequences $\{\xi_{\alpha0}\}$.} of $K_{2}$ imply stability for a cold, uniform density plasma. Equation
(\ref{eq:EM-2}) shows, by $G_{2}(0)=0$, that the force at $\xi_{\alpha}$
due to $\xi_{\beta}$ with $\beta=\alpha$, the self-force, is zero.
Notice, however, that the term in Eq.~(\ref{eq:PertEoM-3}), with
$\alpha=\beta$, is not zero, i.e.~$K_{2}(0)=\int K(y)^{2}dy$ does
not vanish.

Taking the limit $q=m=1/N_{p}\to0$ in Eq.~(\ref{eq:PertEoM-3}),
we find
\begin{equation}
\delta\ddot{\xi}(x,t)=-\int K_{2}(x-y)\delta\xi(y,t)dy.\label{eq:MF-EOM-linear}
\end{equation}
In Fourier space this becomes
\begin{equation}
\omega^{2}\delta\tilde{\xi}(k,\omega)=\tilde{K}_{2}(k)\delta\tilde{\xi}(k,\omega)\label{eq:Cold-DispReln}
\end{equation}
or $\omega^{2}=\tilde{K}_{2}(k)$, where $k=2\pi m$, and the width
of $\tilde{K}_{2}(k)$ is $1/w$. For $w\to0$ this gives the plasma
oscillation $\omega^{2}=\omega_{p}^{2}=1$, and the non-negative $\tilde{K}_{2}(k)$
factor gives the modification due to finite kernel width, introducing
dispersion into the cold plasma oscillation frequency.

\subsection{Linearization for a cold beam}
\label{subsec:Linearization-for-a}

We start with a cold drifting beam, with the equilibrium distribution
\begin{equation}
F_{0}(v)=\delta(v-v_{0}).\label{eq:F0-for-cold-beam}
\end{equation}
For this equilibrium we take
\begin{equation}
\xi_{\alpha0}(t)=\xi_{\alpha0}(0)+v_{0}t,\label{eq:Beam-displacements}
\end{equation}
where $\xi_{\alpha0}(0)$ are $N_{p}$ initial points on the lattice
introduced earlier. This leads to
\begin{equation}
\rho_{0}(x,t)=q\sum_{\beta}K(x-\xi_{\beta0}(0)-v_{0}t).\label{eq:EquilDensity}
\end{equation}
This unperturbed density is again uniform, $\rho_{0}(x,t)=1$, by
the POU property on the lattice, for any $v_{0}$. The equilibrium
electric field, satisfying Gauss's law is, from Eq.~(\ref{eq:Beam-displacements}),
\begin{equation}
E_{0}(x,t)=q\sum_{\beta}G(x-\xi_{\beta0}(0)-v_{0}t).\label{eq:Equil-E-field}
\end{equation}
This is zero by the POU property on the lattice, so that this unperturbed
state is a uniform density, zero electric field equilibrium.

The perturbations of the positions in Eq.~(\ref{eq:EM-1}) and (\ref{eq:EM-2})
are done as in the previous section. The force is given in Eq.~(\ref{eq:Force-at-beta-2})
and its perturbation is 
\begin{equation}
\delta F_{\alpha}=-q^{2}\sum_{\beta}G_{2}'(\xi_{\alpha0}(0)-\xi_{\beta0}(0))\left(\delta\xi_{\alpha}-\delta\xi_{\beta}\right),\label{eq:ColdBeamPertForce0}
\end{equation}
where the $v_{0}t$ terms have canceled. As in Sec.~\ref{subsec:LinearizationForaCold},
we conclude that for $\delta\boldsymbol{\xi}$ orthogonal to the subspace
spanned by $\boldsymbol{c}_{p}$, this satisfies
\begin{equation}
\delta\ddot{\xi}_{\alpha}=-q\sum_{\beta}K_{2}(\xi_{\alpha0}-\xi_{\beta0})\delta\xi_{\beta}.\label{eq:ColdBeamPertForce}
\end{equation}

The perturbed equations of motion are, therefore, $(d^{2}/dt^{2})\xi_{\alpha}=\delta F_{\alpha}/m$,
where $d/dt$ is in the frame moving at velocity $v_{0}$, $d/dt=(\,\dot{}\,)=\partial_{t}+v_{0}\partial_{x}$.
Taking the limit $q=\Delta_{p}\to0$ and recalling $m=q$, this leads
to
\begin{equation}
\delta\ddot{\xi}(x,t)=-\int dyK_{2}(x-y)\delta\xi(y,t).\label{eq:Cont-xEoM}
\end{equation}
In the Fourier series representation with period unity this becomes
\begin{equation}
(\omega-kv_{0})^{2}\delta\tilde{\xi}(k,\omega)=\tilde{K}_{2}(k)\delta\tilde{\xi}(k,\omega)\label{eq:FourierEoM}
\end{equation}
with $k=2\pi m$, leading to the dispersion relation
\begin{equation}
(\omega-kv_{0})^{2}=\tilde{K}_{2}(k).\label{eq:ColdBeamDR}
\end{equation}
The symmetric, positive definite property of $K_{2}$ implies $\tilde{K}_{2}(k)\geq0,\,\,\tilde{K}_{2}(-k)=\tilde{K}_{2}(k)$.
Equation (\ref{eq:ColdBeamDR}) is the cold beam plasma oscillation
dispersion relation for velocity $v_{0}$, taking the dispersive particle
shape into account with the factor $\tilde{K}_{2}(k)$. Again, we
have, with $K_{2}(x)=K_{20}(x/w)/w$ and thus $\tilde{K}_{2}(k)=\tilde{K}_{20}(kw)$.
With $w\to0$ we have $(\omega-kv_{0})^{2}=\tilde{K}_{20}(kw)\to\omega_{p}^{2}=1$.
A similar analysis for non-zero temperatures gives, to lowest order, the correct $k^2\lambda_D^2$ term.

\section{Discretized equations of motion approach}
\label{sec:Discrete-equations-of}

In this section we apply our meshfree formalism to a cold non-drifting
plasma, discretized to a grid. We use this to obtain the linearized
equations of motion on a grid. We first will carry out the linearization
procedure in terms of electric field, as done in the momentum conserving
(MCP) discretization\cite{BLBook}. Next, we will turn to a formulation
in terms of the electrostatic potential, as done in an energy conserving
(ECP) discretization\cite{Lewis1970,BLBook,HockneyEastwood,Evstatiev2013,Evstatiev2014,evstatiev_efficient_2023,Dan-Luis}.

\subsection{Discrete equations for momentum conserving PIC}
\label{subsec:Discrete-equations-for-momentum}

We take Eq.~(\ref{eq:Force-at-beta-0}) for the force at $\xi_{\alpha}$
and discretize on a uniform grid $\{x_{i}\}_{i=0}^{N_{g}-1}$, specifically
$x_{i}=i\Delta,$ where $\Delta=1/N_{g}$. We consider the trapezoidal
discretization
\begin{equation}
F_{\alpha}=-q\Delta\sum_{i=0}^{N_{g}-1}E_{i}K(x_{i}-\xi_{\alpha}).\label{eq:BLForce-discrete}
\end{equation}
This is the form that the force takes in momentum conserving PIC (MCP),
in terms of the electric field at the grid points $E_{i}$. For conventional
PIC the kernel width $w$ is a low integer multiple of $\Delta$,
depending on the choice of which spline to use for the kernel. This
implies that the kernel is not well resolved on the grid. For this
reason, the sum in Eq.~(\ref{eq:BLForce-discrete}) does not converge
exactly to the integral in Eq.~(\ref{eq:Force-at-beta-0}) as $\Delta\to0$.
We will discuss this point further in Sec.~\ref{subsec:Trapezoidal-error/aliasing},
and deal with these issues in detail in a forthcoming publication.

Perturbing about an equilibrium with $\rho_{0}(x)=1$ and $E_{0}(x)=0$
as before, we find
\begin{equation}
\delta F_{\alpha}=-q\Delta\sum_{i}\delta E_{i}\,K(x_{i}-\xi_{\alpha0}),\label{eq:BL-Force-discr-linear}
\end{equation}
where $\xi_{\alpha0}$ represents the uniform lattice of equilibrium
particle positions. The term in which $K(x_{i}-\xi_{\alpha})$ is
perturbed to give $-K'(x_{i}-\xi_{\alpha0})\delta\xi_{\alpha}$ is
proportional to the equilibrium $E_{i0}$, which is zero. The perturbed
electric field is computed from the perturbed electron density by
Gauss's law on a grid. (As we discuss below, this is valid whether
or not the scalar potential $\phi$ is used.)

We now use Eq.~(\ref{eq:E(x)-from-G0}), where $\rho$ is the electron
density, satisfying $\int\rho(y,t)dy=1$. The function $G_{0}$ is
the Green's function for Gauss's law, which was introduced in Sec.~\ref{subsec:Meshfree-equations-of},
the field due to charge density $1-\delta(x)$ with zero potential
difference. As in Sec.~\ref{subsec:Meshfree-equations-of}, we have
Eqs.~(\ref{eq:ODE-for-GF-G0}) and (\ref{eq:G0-Greens-function})
for $-1/2<x<1/2$; $G_{0}$ is extended to be periodic with period
$1$. (As shown in Sec.~\ref{subsec:Meshfree-equations-of}, $G$
defined in Eq.~(\ref{eq:G=00003DG0,smoothed}) satisfies Eq.~(\ref{eq:G'=00003D1-K})).)
The function $G_{0}(x)$ is the limit as $w\to0$ of $G(x)$. From
these conditions and Eq.~(\ref{eq:E(x)-from-G0}) we easily find
\begin{equation}
\partial_{x}E(x)=1-\rho(x),\label{eq:Gauss'-Law}
\end{equation}
i.e.~Gauss's law. 

One approach to discretizing Eq.~(\ref{eq:E(x)-from-G0}) is to take
\begin{equation}
E_{i}=\Delta\sum_{j}G_{0}(x_{i}-x_{j})\rho_{j}.\label{eq:E-from-G0-discrete}
\end{equation}
The centered difference equation satisfied by $E_{i}$ in Eq.~(\ref{eq:E-from-G0-discrete})
involves $G_{0}$, with $\sigma(0)=0$ and noting $\Delta\sum_{j}\rho_{j}=1$:
\[
\frac{E_{j+1}-E_{j-1}}{2\Delta}=\frac{1}{2}\sum_{k}\left(G_{0}(x_{j+1}-x_{k})-G_{0}(x_{j-1}-x_{k})\right)\rho_{j}.
\]
A few calculations show
\begin{equation}
\frac{E_{j+1}-E_{j-1}}{2\Delta}=1-\frac{\rho_{j-1}+2\rho_{j}+\rho_{j+1}}{4}.\label{eq:Gauss-difference-1}
\end{equation}
This particular discretization of Gauss's law is second order accurate
with a circulant mass matrix with $R_{ij}=r_{i-j}$, with rows $(\dots,0,1/4,1/2,1/4,0,\dots)$),
applied to the density. Alternately, we can compute the right--sided
finite difference centered at $j+1/2$, obtaining
\begin{equation}
\frac{E_{j+1}-E_{j}}{\Delta}=1-\frac{\rho_{j}+\rho_{j+1}}{2}.\label{eq:Gauss-difference-2}
\end{equation}
This is a trapezoidal integrator of Gauss's law, also second order
accurate, but with a simpler mass matrix. The representation in Eq.~(\ref{eq:E-from-G0-discrete})
is periodic, valid for both forms, Eq.~(\ref{eq:Gauss-difference-1})
and Eq.~(\ref{eq:Gauss-difference-2}).

The standard way of computing the electric field\cite{BLBook} is
to solve $-(\phi_{j+1}-2\phi_{j}+\phi_{j-1})/\Delta^{2}=1-\rho_{j}$
for $\phi_{i}$ and write $E_{j}=-(\phi_{j+1}-\phi_{j-1})/2\Delta$.
By performing DFT analysis we find that the discrete Gauss's law forms
of the standard approach and the two approaches in Eqs.~(\ref{eq:Gauss-difference-1})
and (\ref{eq:Gauss-difference-2}) are identical, giving $\tilde{E}_{k}=\left(i\Delta\cot(k\Delta/2)/2\right)\tilde{\rho}_{k}$.

Perturbing Eq.~(\ref{eq:E-from-G0-discrete}) we find
\begin{equation}
\delta E_{i}=\Delta\sum_{j}G_{0}(x_{i}-x_{j})\delta\rho_{j}=\Delta\sum_{j}G_{0,i-j}\delta\rho_{j}.\label{eq:Pert-E-from-G0}
\end{equation}
Note the antisymmetry and the circulant nature of the matrix $G_{0,i-j}=G_{0}(x_{i}-x_{j})$
(mod$(1)$).

Using Eq.~(\ref{eq:KDE}) we find
\begin{equation}
\delta\rho_{j}=-q\sum_{\beta}K'(x_{j}-\xi_{\beta0})\delta\xi_{\beta}.\label{eq:Pert-rho-KDE}
\end{equation}
Substituting Eqs.~(\ref{eq:Pert-E-from-G0}) and (\ref{eq:Pert-rho-KDE})
into Eq.~(\ref{eq:BL-Force-discr-linear}) we find
\begin{equation}
\delta\ddot{\xi}_{\alpha}=q\Delta^{2}\sum_{ij\beta}K(x_{i}-\xi_{\alpha0})G_{0,i-j}K'(x_{j}-\xi_{\beta0})\delta\xi_{\beta}\label{eq:Pert-force-BL-1}
\end{equation}
or
\begin{equation}
\delta\ddot{\xi}_{\alpha}=-q\sum_{\beta}M_{\alpha\beta}\delta\xi_{\beta};\label{eq:Pert-Force-BL-2}
\end{equation}
\begin{equation}
M_{\alpha\beta}=-\Delta^{2}\sum_{ij}K(x_{i}-\xi_{\alpha0})G_{0,i-j}K'(x_{j}-\xi_{\beta0}).\label{eq:Dyn-Mx-BL-1}
\end{equation}
Partition of unity leads to $\sum_{\beta}K'(x_{j}-\xi_{\beta0})c_{p,\beta}=0$
for $\boldsymbol{c}_{p}=(1,1,\dots,1)_{N_{p}}^{T}$ and we therefore
have $\sum_{\beta}M_{\alpha\beta}c_{p,\beta}=0$, so that $\delta\boldsymbol{\xi}\perp\boldsymbol{c}$
can be assumed in Eq.~(\ref{eq:Pert-Force-BL-2}). This defines the
\emph{dynamical matrix} $M$, which is of the form $\propto-K^{T}G_{0}K'$
(for MCP) and is not manifestly symmetric or positive definite, preventing
an immediate conclusion on stability. The eigenvalues of $qM$, $\lambda=\omega^{2}$,
must be computed directly. 

As an aside, the integral form is
\begin{equation}
M_{\alpha\beta}=-\int dxdyK(x-\xi_{\alpha0})G_{0}(x-y)K'(y-\xi_{\beta0}),\label{eq:M^=00007Bmc=00007D_Integral}
\end{equation}
again with the caveat that, with $\Delta\to0$ and with $w\propto\Delta$,
the kernel is not resolved accurately, as discussed in Sec.~\ref{subsec:Trapezoidal-error/aliasing}
and Appendix \ref{app:Trapezoidal}. Nevertheless, integrating Eq.~(\ref{eq:M^=00007Bmc=00007D_Integral})
by parts we find
\begin{equation}
M_{\alpha\beta}=-\int dxdyK(x-\xi_{\alpha0})G_{0}'(x-y)K(y-\xi_{\beta0}),\label{eq:M-matrix;Delta=00003D0}
\end{equation}
and using Eq.~(\ref{eq:ODE-for-GF-G0}) we find
\begin{equation}
M_{\alpha\beta}=\int dxK(x-\xi_{\alpha0})K(x-\xi_{\beta0})-\int dxK(x-\xi_{\alpha0})\,\int dyK(y-\xi_{\beta0}).\label{eq:M-mat;Delta=00003D0-2}
\end{equation}
Using Eqs.~(\ref{eq:q=00003D,intK=00003D1}) and (\ref{eq:K2-2-forms})
we find, in this limit,
\begin{equation}
M_{\alpha\beta}\to K_{2}(\xi_{\alpha0}-\xi_{\beta0})-1=K_{2\alpha\beta}-1.\label{eq:M=00003D1-K-BL}
\end{equation}

The proof of momentum conservation in the MCP formulation\cite{BLBook}
is based on the antisymmetry of the discretization of Gauss's law,
equivalent to the antisymmetry of $G_{0,i-j}$ noted in Eq.~(\ref{eq:Pert-E-from-G0}).
See Sec.~\ref{subsec:MCP-discretization}. Note that this asymmetry
does not lead to any evident properties of the matrix $M_{\alpha\beta}$
in Eq.~(\ref{eq:Dyn-Mx-BL-1}), such as being symmetric positive
definite (SPD).

Direct discretization of Eq.~(\ref{eq:M-mat;Delta=00003D0-2}) leads
to
\[
M_{\alpha\beta}=\Delta\sum_{ij}K(x_{i}-\xi_{\alpha0})K(x_{j}-\xi_{\beta0})-1.
\]

This matrix, of the form $K^{T}K$, is manifestly SPD, and has sparseness
because of the compact support of the kernel $K$. However, this matrix,
representing another distinct discretization, is not identical to
that in the MCP discretization, or in the ECP discretization in the
next subsection.

\subsection{Discrete equations for energy conserving PIC}
\label{subsec:Discrete-equations-for}

We begin again with Eq.~(\ref{eq:Force-at-beta-0}), but in contrast
with the last section, we write the force in terms of the electrostatic
potential as $F_{\alpha}=q\int\phi'(x)K(x-\xi_{\alpha})$ and integrate
by parts to obtain
\begin{equation}
F_{\alpha}=-q\int\phi(x)K'(x-\xi_{\alpha})dx\label{eq:Force-w-phi}
\end{equation}
and discretize it on the grid as
\begin{equation}
F_{\alpha}=-\Delta q\sum_{i}\phi_{i}K'(x_{i}-\xi_{\alpha}).\label{eq:Discr-force-w-phi}
\end{equation}
This representation for the force in terms of $\phi_{i}$ occurs in
the energy conserving or variational PIC formulation (ECP).\cite{BLBook,Lewis1970,Evstatiev2013,Evstatiev2014}.
The Poisson equation $-\phi''(x)=1-\rho(x)$ leads to $\phi(x)=\int L(x-y)\rho(y)dy$
with $-L''(x)=1-\delta(x)$, i.e.~$-L''(x)=G_{0}'(x)$. This yields
$L(x)=(|x|-x^{2})/2$, so that $L$ is the potential of a single point
electron with its share of neutralizing ion charge. That is, $L(x)$
is the Green's function for the Poisson equation with periodic boundary
conditions on $\phi$ and a neutralizing ion background. Linearizing
Eq.~(\ref{eq:Discr-force-w-phi}), we find
\begin{equation}
\delta F_{\alpha}=-\Delta q\sum_{i}\delta\phi_{i}K'(x_{i}-\xi_{\alpha0}).\label{eq:Pert-force-w-phi}
\end{equation}
The term in which $K'$ is varied is zero because the equilibrium
potential $\phi_{i0}$ is constant and $\sum_{i}K''(x_{i}-\xi_{\alpha0})$
is zero by POU on the grid, where $\Delta$ is assumed to be an integer
times $q$ (recalling $N_{p}=N_{ppc}N_{g}$.) The perturbation $\delta\phi$
is found by the Poisson equation $\delta\phi''=\delta\rho$, and indeed
gives $\delta\phi(x)=\int L(x-y)\delta\rho(y)dy$. The discrete form
is $\delta\phi_{i}=\Delta\sum_{j}L_{i-j}\delta\rho_{j}$, leading
to
\begin{equation}
\delta\phi_{i}=\Delta\sum_{j}L_{i-j}\delta\rho_{j};\label{eq:Lin-rho->phi}
\end{equation}
$-L_{i-j}$ is circulant and positive definite (with a null space
$\delta\boldsymbol{x}\propto\boldsymbol{c}_{g}=(1,1,\dots,1)_{N_{g}}$.)
This discrete form is related to the discrete perturbed Poisson equation,
i.e.~
\begin{equation}
\delta\rho_{i}=\Delta^{-1}\sum_{j}\kappa_{i-j}\delta\phi_{j},\label{eq:Lin-phi->rho}
\end{equation}
with $L=\kappa{}^{-1}$. These operators are symmetric and negative
definite, for example with rows of $\kappa$ equal to $\Delta^{-2}(0,\dots,1,-2,1,0,\dots,0)_{N_{g}}$.
Substituting as in the last subsection we find $m\delta\ddot{\xi}=\delta F_{\alpha}$
with $m=q$ and

\begin{equation}
\delta\ddot{\xi}_{\alpha}=-q\sum_{\beta}M_{\alpha\beta}\delta\xi_{\beta},\label{eq:Pert-force-ECP}
\end{equation}
\begin{equation}
M_{\alpha\beta}=-\Delta^{2}\sum_{ij}K'(x_{i}-\xi_{\alpha0})L_{i-j}K'(x_{j}-\xi_{\beta0}).\label{eq:Mab-ECP}
\end{equation}
The ECP dynamical matrix $M$ is of the form $\propto-K'^{T}LK'$
with $-L$ SPD, so that, unlike in the MCP case, the dynamical matrix
$M$ is SPD. Note the relation $\sum_{\beta}K'(x_{j}-\xi_{\beta0})=0$,
i.e.~the constant vector $\boldsymbol{c}_{p}=(1,1,\dots,1)_{N_{p}}$
is again in the null space of $M$. 

The integral form of Eq.~(\ref{eq:Mab-ECP}) (while the sum is not
equal to the integral as $\Delta\to0$) is
\begin{equation}
M_{\alpha\beta}=-\int dxdyK'(x-\xi_{\alpha0})L(x-y)K'(y-\xi_{\beta0})\label{eq:Integral-form-MCP-1}
\end{equation}
or
\begin{equation}
M_{\alpha\beta}=\int dxdyK(x-\xi_{\alpha0})L''(x-y)K(y-\xi_{\beta0})\label{eq:Integral-form-MCP-2}
\end{equation}
with $L''(x)=\delta(x)-1$ or
\[
M_{\alpha\beta}=\int dxdyK(x-\xi_{\alpha0})K(x-\xi_{\beta0})-\int dxK(x-\xi_{\alpha0})\,\times\,\int dyK(y-\xi_{\beta0})
\]
or
\[
M_{\alpha\beta}=K_{2}(\xi_{\alpha0}-\xi_{\beta0})-1.
\]
This is the same form $M=K_{2}-\boldsymbol{c}_{p}\boldsymbol{c}_{p}^{T}$
as in the integral form in the last subsection.

Returning to the sum for $\Delta>0$, we can write, using Eq.~(\ref{eq:Pert-force-ECP}),
\[
\omega^{2}\delta\xi_{\alpha}=q\sum_{\beta}M_{\alpha\beta}\delta\xi_{\beta}.
\]
Note that in this ECP cold plasma case, the dynamical matrix $M$
is SPD (but not strictly positive definite), directly implying stability.
The eigenvalues of $qM$, $\lambda=\omega^{2}$ are all real and non-negative.

\subsection{Conservation laws for the linearized equations}
\label{subsec:Conservation-laws-for}

We can again write the perturbed total momentum as 
\[
\delta P=q\sum_{\alpha}\delta u_{\alpha}
\]
(with $m=q$) leading to
\begin{equation}
\delta\dot{P}=\sum_{\alpha}\delta F_{\alpha}=-q^{2}\sum_{\alpha\beta}M_{\alpha\beta}\delta\xi_{\beta}.\label{eq:P-dot}
\end{equation}

\subsubsection{ECP discretization}
\label{subsec:ECP-discretization}

For the ECP case, the expression for $\dot{P}$ can be written as
\begin{equation}
\dot{P}=-q^{2}(\boldsymbol{c}_{p},M\delta\boldsymbol{\xi})=-q^{2}(M\boldsymbol{c}_{p},\delta\boldsymbol{\xi}),\label{eq:P-dot-ECP}
\end{equation}
where we have used symmetry of $M$. We have noted $M\boldsymbol{c}_{p}=0$
in the last section, so conservation of momentum $\dot{P}=0$ follows. 

Because of the form of the equations of motion for the ECP case, $\delta\dot{\boldsymbol{\xi}}=\delta\boldsymbol{u}$,
$\delta\boldsymbol{\dot{u}}=-qM\delta\boldsymbol{\xi}$ with $M$
symmetric, these are linear Hamiltonian equations in canonical form,
and for infinitesimal time step, energy is conserved.

\subsubsection{MCP discretization}
\label{subsec:MCP-discretization}

For the MCP case, with $M\propto-K^{T}G_{0}K'$ we have
\begin{equation}
\dot{P}=-q^{2}(M^{T}\boldsymbol{c}_{p},\delta\boldsymbol{\xi}).\label{eq:dot=00007BP=00007D-MCP}
\end{equation}
We have $M^{T}\boldsymbol{c}_{p}=-K'^{T}G_{0}^{T}K\boldsymbol{c}_{p}=K'^{T}G_{0}K\boldsymbol{c}_{p}$.
The factor $K\boldsymbol{c}_{p}$ is not zero, but we do have $K\boldsymbol{c}_{p}\propto\boldsymbol{c}_{g}$,
and the antisymmetry of $G_{0}$ gives $G_{0}\boldsymbol{c}_{g}=0$.
We conclude that momentum is conserved, and that this property depends
on the antisymmetry of $G_{0}$, i.e.~antisymmetry of the relation
between $\rho$ and $E$. This property was noted in Ref.~\cite{BLBook,HockneyEastwood},
and discussed in Sec.~\ref{subsec:Discrete-equations-for-momentum}.

On the other hand, energy is not conserved because of the lack of
symmetry in $M$ for the MCP case. That is, the linearized equations
in this case are not Hamiltonian in canonical variables.

\section{Properties of the matrix $M$ for MCP and ECP}
\label{sec:Properties-of-the}

In this section we explore properties of the matrices $M$ in the
MCP and ECP discretizations. The first of these properties is the
trapezoidal rule error in the discretized force on macroparticles,
i.e.~the occurrence of aliasing. We also discuss the block circulant
nature of these $N_{p}\times N_{p}$ dynamical matrices, and derive
a dimensionally reduced dynamical matrix on the grid ($N_{g}\times N_{g}$).
We note that this matrix is circulant, and discuss consequences of
the fact that these circulant matrices are normal.

\subsection{Trapezoidal error and aliasing}
\label{subsec:Trapezoidal-error/aliasing}

We first consider the error in taking the meshfree particle force
in Eq.~(\ref{eq:Force-at-beta-0}) and its perturbed form in Eq.~(\ref{eq:PertEoM-3})
and replacing with the discretized trapezoidal rule form of the latter.
The resulting equation of motion is shown in Eqs.~(\ref{eq:Pert-force-BL-1}).

First, we recall that the integral form of the matrix $M$ in the
MCP discretization is the meshfree form in Eq.~(\ref{eq:M^=00007Bmc=00007D_Integral}).(These
properties hold for both the MCP and ECP discretization, but are more
important for the former because the dynamical matrix is SPD for the
latter.) Now consider the shift $\xi_{\alpha0}\to\xi_{\alpha0}+\epsilon$,
to be discussed in more detail in Sec.~\ref{sec:Shifted-lattice}.
It is straightforward to show, by the substitution $x'=x-\epsilon,\,\,y'=y-\epsilon$,
\begin{equation}
M_{\alpha\beta}(\xi_{\alpha0}+\epsilon,\xi_{\beta0}+\epsilon)=-\int dxdyK(x'-\xi_{\alpha0})G_{0}(x'-y')K'(x'-\xi_{\beta0}).\label{eq:Triple-M-delta0-2}
\end{equation}
That is, the shifted form of the matrix elements are independent of
$\epsilon$. The same property holds for the matrix in the ECP form
in its integral form.

As mentioned in Sec.~\ref{sec:Discrete-equations-of}, there is one
caveat about the limit as $\Delta\to0$, related to the fact that,
in the conventional PIC schemes, the particle widths (kernel widths)
are proportional to $\Delta$, with a $O(1)$ proportionality factor
($2\text{ to }4$ for linear to cubic), i.e.~depending on the particle
shape. That is, the kernel (or its derivative) is not well resolved,
even for small $\Delta$. This point suggests that the limit $\Delta\to0$
may exist, but it will not necessarily be equal to the meshfree integral
form that holds for $\Delta\to0$ with the kernel width $w$ fixed.
This issue is discussed in more detail in Appendix \ref{app:Trapezoidal}.

Returning to the discrete form for $\Delta>0$, the error in computing
the force by its discretized approximation, the trapezoidal rule formula,
is first analyzed by expanding the integrand in a Taylor series about
each grid point and summing these. For integrands with a jump in the
$n$'th derivative, the error is $O(\Delta^{n+1})$. For details,
see Appendix \ref{app:Trapezoidal}.

An alternate approach to estimating this error, also discussed in
detail in Appendix \ref{app:Trapezoidal}, is to compute the Fourier
series for the integrand, assuming periodicity of period $1$. Then
the discrete Fourier transform (DFT) coefficient is the sum of the
aliases, namely
\begin{equation}
b_{m}=\sum_{p=-\infty}^{\infty}a_{m+pN_{g}},\label{eq:FourierSeries->DFT}
\end{equation}
where $a_{m}$ are the Fourier series coefficients and $b_{m}$ are
the DFT values. Specializing to $m=0$, where $b_{0}$ is the trapezoidal
rule approximation to the integral, gives the aliasing error (or \emph{alias})
$A_{0}$
\begin{equation}
A_{0}\doteq b_{0}-a_{0}=\sum_{p\neq0}a_{pN_{g}}.\label{eq:b0-a0Alias-1}
\end{equation}

For an integrand with a jump discontinuity, we have $a_{m}\sim1/m$
as $m\to\infty$, which, with $m=N_{g}p$, leads to
\begin{equation}
A_{0}\propto\frac{2}{N_{g}}\sum_{p=1}^{\infty}(\cdots)\frac{1}{p}.\label{eq:b0-a0Alias-10}
\end{equation}
Typically the coefficients $(\cdots)$ are alternating in sign, leading
to convergence and the conclusion 
\begin{equation}
A_{0}\propto\Delta.\label{eq:A0ProptoDelta}
\end{equation}
(Subtleties related to the order of the terms in the sum are discussed
in Appendix \ref{app:Trapezoidal}.)

For a continuous integrand with a jump in derivative, we have $a_{m}\sim1/m^{2}$
as $m\to\infty$, leading to the estimate in the aliasing error
\begin{equation}
A_{0}\propto\frac{2}{N_{g}^{2}}\sum_{p=1}^{\infty}\frac{1}{p^{2}}\approx\frac{2}{N_{g}^{2}}\int_{1}^{\infty}\frac{dp}{p^{2}}\propto\frac{1}{N_{g}^{2}}=\Delta^{2},\label{eq:b0-a0Alias-2}
\end{equation}
in agreement with the trapezoidal error above, and similar considerations
imply that for a jump in the $n$'th derivative, the aliasing error
has
\begin{equation}
A_{0}\propto\Delta^{n+1}.\label{eq:b0-a0Alias3}
\end{equation}

In Appendix \ref{app:Trapezoidal}, it is shown that for MCP and for
either the linear or the quadratic spline kernel, the matrix elements
$M_{\alpha\beta}$ have aliasing errors due to discontinuities of
the form
\begin{equation}
A_{0}=b_{0}-a_{0}=C_{1}\Delta\label{eq:Tent-step-aliasing}
\end{equation}
for small $\Delta$, for some constant $C_{1}$. ($C_{1}$ may depend
on $N_{ppc}$.)

In addition to the aliasing due to discontinuities, there is a
separate contribution to the aliasing, one due to the small width
$w$ of the kernel $K$, also discussed in Appendix \ref{app:Trapezoidal}.
To estimate the order of magnitude of the aliasing due to $w$, we
estimate the number of aliases, i.e.~the number of terms in Eq.~(\ref{eq:b0-a0Alias-1})
that contribute appreciably to the DFT sum. The width of the kernel
in Fourier space scales as $m_{w}\sim1/w\sim1/\Delta\sim N_{g}$,
and the Nyquist mode has $m_{N}=N_{g}/2.$ This leads to the number
of aliases, equal to $m_{w}/m_{N}\sim1$, i.e.~converging to a $O(1)$
constant as $\Delta\to0$. This leads to the aliasing due to the kernel
width scaling as 
\[
A_{0}=C_{0}
\]
for $N_{ppc}$ fixed, i.e.~$N_{ppc}=\Delta/q$ fixed. (This is due
to the issue brought up above: because $w\propto\Delta$, the kernel
is not well resolved and the limit $N_{g}\to\infty$ ($\Delta\to0$)
is not zero. For only the lowest order terms in $\Delta$, we can
combine with Eq.~(\ref{eq:Tent-step-aliasing}) to find, for MCP
and either the linear kernel or the quadratic spline, the aliasing
error
\begin{equation}
E(\Delta)=A_{0}=C_{0}+C_{1}\Delta.\label{eq:C0+C1=00005CDelta-fla}
\end{equation}

As in Appendix \ref{app:Trapezoidal}, for the quadratic spline kernel
we have the same expression, possibly with different coefficients
$C_{0},\,\,C_{1}$. As we have discussed, we expect $C_{0}>0$, so
that the aliasing error does not converge to zero as $\Delta\to0,$
i.e.~the sum may converge but does not converge to the integral appearing
in the meshfree formulation in Sec.~\ref{sec:Meshfree-formulation},
e.g.~Eq.~(\ref{eq:K2-2-forms}). This is because the limit in conventional
PIC is taken with $w\propto\Delta$. That is, the summand of Eq.~(\ref{eq:Dyn-Mx-BL-1})
(or Eq.~(\ref{eq:Mab-ECP})) is never well resolved by the grid.
A related aspect of PIC with $w\propto\Delta$ is that, as higher
order splines are used, the kernel becomes smoother and also its width
increases, both contributing to a decrease in the aliasing.

These arguments apply equally well for the MCP case and the ECP case.
In Sec.~\ref{sec:Shifted-lattice} we present results related to
the scaling of $A_{0}$ with respect to $q/\Delta=N_{ppc}$.

\subsection{Block circulant nature of $M$ for ECP and MCP}
\label{sec:Block-circulant}

For application of the MCP and ECP results of Sections \ref{subsec:Discrete-equations-for-momentum}
and \ref{subsec:Discrete-equations-for}, we start by requiring $N_{p}\geq N_{g}$
and further $N_{p}=N_{ppc}N_{g}$ with $N_{ppc}$, the number of particles
per cell, a positive integer, so we have $\Delta=N_{ppc}q$. Specializing
for the moment to the ECP case, note that for the matrix in Eq.~(\ref{eq:Mab-ECP})
we have
\[
M_{\alpha+N_{ppc},\beta+N_{ppc}}=-\Delta^{2}\sum_{ij}K'(x_{i}-\xi_{\alpha0}-N_{ppc}q)L_{i-j}K'(x_{j}-\xi_{\beta0}-N_{ppc}q)
\]
\begin{equation}
=-\Delta^{2}\sum_{ij}K'(x_{i}-\Delta-\xi_{\alpha0})L_{i-j}K'(x_{j}-\Delta-\xi_{\beta0}),\label{eq:BC-proof}
\end{equation}
where $\Delta=N_{ppc}q$ has been used. Using $x_{i}-\Delta=x_{i-1}$
(mod$(1)$) (evenly spaced points on the period), and the fact that
$L$ is circulant, we conclude
\begin{equation}
M_{\alpha+N_{ppc},\beta+N_{ppc}}=M_{\alpha\beta}.\label{eq:BC-Result}
\end{equation}
That is, $M$ is a \emph{block-circulant} $N_{p}\times N_{p}$ matrix,
with $N_{ppc}\times N_{ppc}$ blocks. It is straightforward to show
that this block-circulant property holds for the MCP case of Eq.~(\ref{eq:Dyn-Mx-BL-1})
as well. (In the special case $N_{ppc}=1$ the matrices are \emph{circulant},
i.e.~we have $M_{\alpha+1,\beta+1}=M_{\alpha\beta}$ or $M_{\alpha\beta}=M_{\alpha-\beta}$.)
The eigenvalues and eigenvectors of a circulant matrix can be found
by means of the discrete Fourier transform (DFT), as is well known\cite{richtmyer1967difference,Press2007}. 

There are methods that enable the fast computation of eigenvalues
of block-circulant matrices, based on a generalization of the DFT
methods used for circulant matrices\cite{gray2006toeplitz}.

\subsection{Dimensional reduction of $M$ to a grid-sized matrix}
\label{subsec:Dimensional-reduction-of}

As noted in the previous subsection, the dynamical matrix $M$ of
in both the MCP form and the ECP form, is block-circulant, aiding
in computation. This matrix is dimensioned $N_{p}\times N_{p}$ but
with rank at most $N_{g}\ll N_{p}$, allowing for a null space of
dimension at least $N_{p}-N_{g}$. We will call the associated zero
eigenvalues the \emph{trivial eigenvalues}. In Appendix \ref{app:DimReduction}
we discuss a method of dimensional reduction, in which a grid-sized
(dimensioned $N_{g}\times N_{g}$) matrix $\hat{M}$, is found. The
nontrivial eigenvalues of $M$ and the eigenvalues of $\hat{M}$ are
identical. In the MCP form the dynamical matrix is of the form 
\begin{equation}
M\propto-K^{T}G_{0}K',\label{eq:M-matrix-form}
\end{equation}
and the approach of Appendix \ref{app:DimReduction} introduces the
dimensional reduction to the $N_{g}\times N_{g}$ matrix
\begin{equation}
\hat{M}\propto-K'(K^{T}G_{0}).\label{eq:N-matrix-form}
\end{equation}
(The reduction $K^{T}G_{0}K'\to(G_{0}K')K^{T}$, also $N_{g}\times N_{g}$,
works just as well.)

The dimensionally reduced $N_{g}\times N_{g}$ matrix $\hat{M}$ has
several important properties: (i) As shown in Appendix \ref{app:DimReduction},
its eigenvalues are eigenvalues of $M$, with the trivial eigenvalues
of $M$ (zero eigenvalues of $K^{T}G_{0}$) missing; (ii) It is circulant,
i.e.~$\hat{M}_{i+1,j+1}=\hat{M}_{ij}$, or $\hat{M}_{ij}=\hat{M}_{i-j}$.
To show the circulant property, we note
\begin{equation}
\hat{M}_{ij}=-\Delta^{2}\sum_{\alpha}K'(x_{i}-\xi_{\alpha0})\sum_{k}K(\xi_{\alpha0}-x_{k})G_{0}(x_{k}-x_{j})\label{eq:Def-Nij}
\end{equation}
and, using $\Delta=N_{ppc}q$,
\[
\hat{M}_{i+1,j+1}=-\Delta^{2}\sum_{\alpha}K'(x_{i}+\Delta-\xi_{\alpha0})\sum_{k}K(\xi_{\alpha0}-x_{k})G_{0}(x_{k}-x_{j+1})
\]
\begin{equation}
=-\Delta^{2}\sum_{\alpha}K'(x_{i}+N_{ppc}q-\xi_{\alpha0})\sum_{l}K(\xi_{\alpha0}-x_{l}-N_{ppc}q)G_{0}(x_{l+1}-x_{j+1})\label{eq:N-circular-pf}
\end{equation}
\[
=-\Delta^{2}\sum_{\alpha}K'(x_{i}-\xi_{\alpha-N_{ppc},0})\sum_{l}K(\xi_{\alpha-N_{ppc},0}-x_{l})G_{0}(x_{l}-x_{j}).
\]
Since the sum over $\alpha$ is over equally spaced lattice points
in a full period, we have
\begin{equation}
\hat{M}_{i+1,j+1}=-\Delta^{2}\sum_{\beta}K'(x_{i}-\xi_{\beta0})\sum_{l}K(\xi_{\beta0}-x_{l})G_{0}(x_{l}-x_{j})=\hat{M}_{ij},\label{eq:N-circular-endpf}
\end{equation}
with $i,\,j$ mod$(N_{g})$. 

The fact that $\hat{M}$ is circulant is related to the fact that
the matrix elements $\hat{M}_{ij}(\epsilon)$, formed by shifting
$\xi_{\alpha0}\to\xi_{\alpha0}+\epsilon$ in Eq.~(\ref{eq:Def-Nij}),
are periodic in $\epsilon$ of period $q$. If we shift the lattice
relative to the grid, we have $\hat{M}_{ij}(\epsilon)$ for $\epsilon=q$,
which is
\[
\hat{M}_{ij}(q)=-\Delta^{2}\sum_{\alpha}K'(x_{i}-\xi_{\alpha0}-q)\sum_{k}K(\xi_{\alpha0}+q-x_{k})G_{0}(x_{k}-x_{j})
\]
\[
=-\Delta^{2}\sum_{\alpha}K'(x_{i}-\xi_{\alpha+1,0})\sum_{k}K(\xi_{\alpha+1,0}-x_{k})G_{0}(x_{k}-x_{j}).
\]
Because we are summing over equally spaced lattice points $\xi_{\alpha,0}$,
we conclude
\begin{equation}
\hat{M}_{ij}(q)=\hat{M}_{ij}(0).\label{eq:Periodicity-of-hat=00007BM=00007D}
\end{equation}

As discussed above, this circulant property implies that the eigenvalues
of $\hat{M}$ can be found easily by DFT analysis. This analysis depends
on the fact that $\hat{M}$ being circulant is equivalent to the property
that $\hat{M}$ commutes with the shift operator $S$, having $(S\boldsymbol{v})_{m}=v_{m+1}$.
This means that eigenvectors of $\hat{M}$ are eigenvectors of $S$,
with components $(\delta x_{m})_{n}=e^{2\pi imx_{n}}=e^{2\pi imn/N_{g}}$.
In particular, if the first row of $\hat{M}$ is $(a_{0},a_{1},\dots,a_{N_{g}-1})$,
the remainder of the rows are found by shifting indices and the eigenvalue
equation is solved by
\begin{equation}
a_{0}+z_{m}a_{1}+z_{m}^{2}a_{2}+\cdots+z_{m}^{N_{g}-1}a_{N_{g}-1}=\lambda a_{0}\label{eq:DFT-eigenvalues}
\end{equation}
for $z_{m}=e^{2\pi imn/N_{g}}$, $N_{g}$ points uniformly spaced
on the unit circle in the complex plane. The constant grid vector
$\boldsymbol{c}_{g}=(1,\cdots,1)_{N_{g}}$ (corresponding to $z_{m}=1$)
is a neutral mode, with eigenvalue zero, and Eq.~(\ref{eq:DFT-eigenvalues})
shows $a_{0}+a_{1}+\cdots+a_{N_{g}-1}=0$. This follows because $G_{0}\boldsymbol{c}_{g}=0$,
because $G_{0}(x)$ is odd in $x$ and $\boldsymbol{c}_{g}$ is even.
For MCP, the same argument holds for the Nyquist ($2\Delta$) mode
$\boldsymbol{n}_{g,j}=(-1)^{j}$. 

For ECP, The uniform mode $\boldsymbol{c}_{g}$ is a neutral mode:
The dimensionally reduced form $K'K'^{T}L$ applied to $\boldsymbol{c}_{g}$
because $L\boldsymbol{c}_{g}=0$. On the other hand, the Nyquist mode
$\boldsymbol{n}_{g}$ does not satisfy $L\boldsymbol{n}_{g}=0$, so
for ECP, the Nyquist mode is not necessarily a neutral mode.

The periodicity of this dimensionally reduced matrix $\hat{M}(\epsilon)$
of period $\epsilon=q$ in Eq.~(\ref{eq:Periodicity-of-hat=00007BM=00007D})
leads with Eq.~(\ref{eq:DFT-eigenvalues}), to periodicity of the
eigenvalues. In Sec.~\ref{subsec:Properties-of-the} we showed that
the eigenvalues of the original dynamical matrix $M$ are periodic
of period $q$ in $\epsilon$, although the matrix elements $M_{\alpha\beta}$
have period $\Delta$.

One final property that we can derive for $\hat{M}$ relates to the
limit $N_{p}\to\infty$. In the MCP case we have $\hat{M}\propto K'K^{T}G_{0}$
or 
\[
\hat{M}_{ij}\propto q\sum_{k}\sum_{\alpha}K'(x_{i}-\xi_{\alpha0})K(\xi_{\alpha0}-x_{k})G_{0}(x_{k}-x_{j}).
\]
For $N_{p}\to\infty$ this approaches
\begin{equation}
\hat{M}_{ij}\propto\sum_{k}\left(\int d\xi K'(x_{i}-\xi)K(\xi-x_{k})\right)G_{0}(x_{k}-x_{j}),\label{eq:hat=00007BM=00007Das-Np=00005Cto=00005Cinfty}
\end{equation}
and as before $\xi\to\xi+\epsilon$ leads to no change. That is, the
aliasing disappears and the growth rate goes to zero as $N_{p}\to\infty$.\textcolor{red}{{}
}The error in the convolution over $\xi$ due to jumps ($(1/m^{2})\cdot(1/m)$)
is $q^{3}$, leading to an error $\sim q^{3}=1/N_{ppc}^{3}$. Because
$w\propto\Delta$ but $w\gg q$ as $N_{p}\to\infty$, the aliasing
of the form $C_{0}$ is absent, leading to 
\[
E(\Delta,q)\propto q^{3}.
\]

\subsection{Non-normal dynamical matrix $M$}
\label{subsec:Non-normal-dynamical-matrix}

Another property of interest is that for the MCP discretization, computation
with several examples shows that the original $N_{p}\times N_{p}$
dynamical matrix $M$ is often non-normal, i.e.~$M^{T}M\neq MM^{T}$,
leading to the possibility of transient growth\cite{Trefethen}. (This
possibility is ruled out for the ECP case, for which $M$ is symmetric
and therefore normal.) On the other hand, since $\hat{M}$ is circulant,
it \emph{is} normal. This is a result of the fact that the eigenvectors,
with $(\delta x_{m})_{n}=e^{2\pi imn/N_{g}}$, are orthogonal. (The
normal property can be shown directly, $\sum_{k}M_{i-k}M_{j-k}=\sum_{l}M_{l-i}M_{l-j}$,
by reindexing.)

As discussed in Appendix \ref{app:DimReduction}, if $M=AB$, with
$A$ of dimension $N_{p}\times N_{g}$ and $B$ of dimension $N_{g}\times N_{p}$,
the eigenvectors associated with nontrivial eigenvalues $\delta\boldsymbol{\xi}$
of $M$ are related to the eigenvectors of $\hat{M}$ by $\delta\boldsymbol{x}=B\delta\boldsymbol{\xi}$.
Also, for MCP we chose $A=K^{T},\,\,B=G_{0}K'$ in the previous subsection.
We conclude that the nontrivial eigenvalues of $M$ are eigenvalues
of $\hat{M}$ and the trivial eigenvalues of $M$ are in the null
space of $B$.

The eigenvectors $\delta\boldsymbol{x}_{m}$ of the circulant (and
therefore normal) matrix $\hat{M}$ are orthonormal, $(\delta\boldsymbol{x}_{m},\delta\boldsymbol{x}_{n})_{N_{g}}=\delta_{mn}$,
where $(\cdot,\cdot)_{N_{g}}$ is the $(l^{2})_{N_{g}}$ or Euclidean
inner product in dimension $N_{g}$. What does this say about the
possible $(l^{2})_{N_{p}}$ orthogonality of the lattice eigenvectors
$\{\delta\boldsymbol{\xi}_{m}\}$, i.e.~the possibility that the
$N_{p}\times N_{p}$ matrix $M$ is normal? We have 
\[
\delta_{mn}=(\delta\boldsymbol{x}_{m},\delta\boldsymbol{x}_{n})_{N_{g}}=(B\delta\boldsymbol{\xi}_{m},B\delta\boldsymbol{\xi}_{n})_{N_{g}},
\]
\begin{equation}
=(\delta\boldsymbol{\xi}_{m},B^{T}B\delta\boldsymbol{\xi}_{n})_{N_{p}},\label{eq:BB-orthogonality}
\end{equation}
where $B^{T}B=K'^{T}G_{0}^{T}G_{0}K'$ is a $N_{p}\times N_{p}$ Gram
matrix. That is, the $\{\delta\boldsymbol{\xi}_{m}\}$ are orthonormal
with respect to the new inner product $\langle\delta\boldsymbol{\xi}|\delta\boldsymbol{\eta}\rangle=(\delta\boldsymbol{\xi},B^{T}B\delta\boldsymbol{\eta})_{N_{p}}$,
i.e.~$\langle\delta\boldsymbol{\xi}_{m}|\delta\boldsymbol{\xi}_{n}\rangle=\delta_{mn}$.
This means that the $\delta\boldsymbol{\xi}$ are orthonormal with
respect to the inner product $\langle\cdot|\cdot\rangle$ but not
the original $(l^{2})_{N_{p}}$ inner product $(\cdot,\cdot)$. Thus,
$M=AB$ is non-normal with respect to $(\cdot,\cdot)$ but normal
with respect to the alternate inner product $\langle\cdot|\cdot\rangle$.
This allows for the possibility that the linearized equations of motion
of $\delta\boldsymbol{\xi}$ can show transient growth when measured
by the norm associated with $(\cdot,\cdot)$, but not when measured
by the norm associated with $\langle\cdot|\cdot\rangle$. For example,
this means that if the imaginary parts of the eigenvalues of $M$
are non-positive, the solution $\delta\boldsymbol{\xi}$ of $\delta\ddot{\boldsymbol{\xi}}=qM\delta\boldsymbol{\xi}$
can initially show growth in $(\delta\boldsymbol{\xi},\delta\boldsymbol{\xi})$
but eventually decay. However, in spite of this decay in the linearized
system, the perturbations might reach sufficient amplitude in the
original nonlinear system that nonlinear effects become important.
We will discuss these issues in a forthcoming publication.

\section{Shifted lattice}
\label{sec:Shifted-lattice}

\begin{figure}
  \centering
  \includegraphics[width=0.5\textwidth]{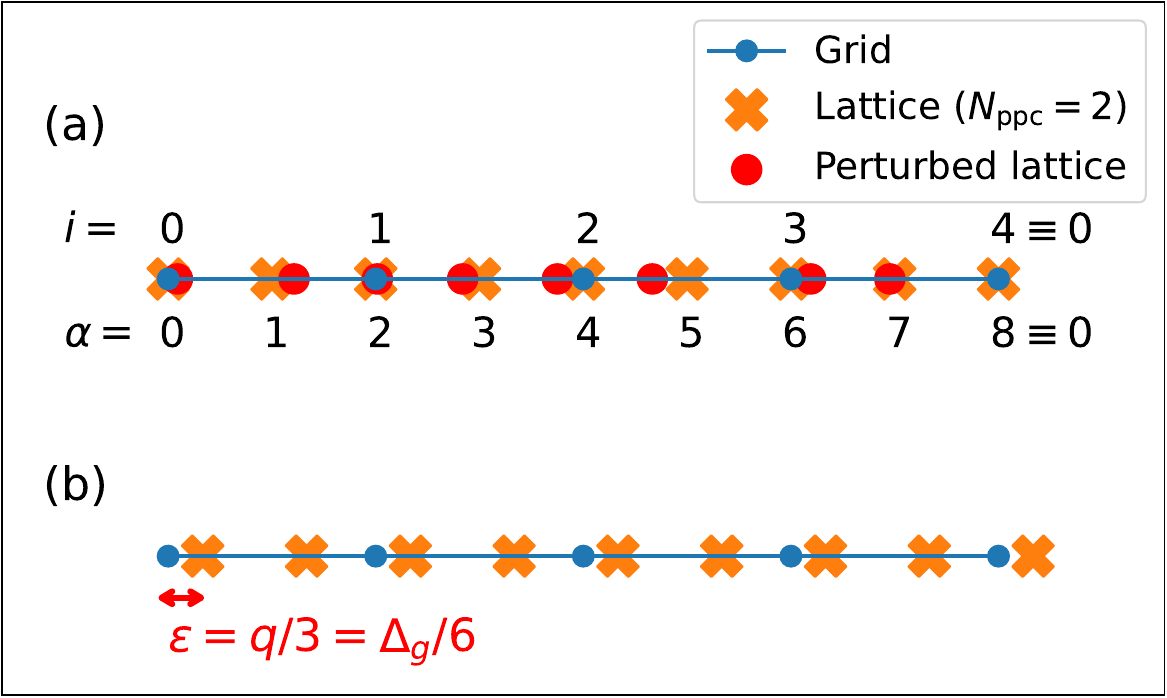}
\caption{Particle lattice points $\xi_{\alpha0}$ (crosses) and grid points
$x_{i}$ (dots) for $N_{g}=4,\,\,N_{p}=8$, with perturbed lattice
points (a). Shifted lattice $\xi_{\alpha0}\to\xi_{\alpha0}+\epsilon$
and grid $x_{i}$ in (b) for $\epsilon=q/3$, without perturbation.}
\label{Fig-lattice-and-grid-2} 
\end{figure}

An obvious way to locate the equilibrium particle lattice $\{\xi_{\alpha0}\}$
relative to the grid $\{x_{i}\}$, with  $N_{ppc}=N_{p}/N_{g}$ a
positive integer, is to set $\xi_{\alpha0}=\alpha q$, with $q=1/N_{p}$
and $x_{i}=i\Delta$, where $\Delta=1/N_{g}$, so we have $\xi_{\alpha=0,0}=x_{i=0}=0$
and $\xi_{\alpha=N_{p},0}=x_{i=N_{g}}=1$, as illustrated in Fig.~\ref{Fig-lattice-and-grid-2}(a)
for $N_{ppc}=2$. In this section we explore the effects of a different
choice of equally spaced lattice, i.e.~of a rigid shift of the lattice,
$\xi_{\alpha0}\to\xi_{\alpha0}+\epsilon$, illustrated in Fig.~\ref{Fig-lattice-and-grid-2}(b).

\subsection{Properties of the dynamical matrix $M$}
\label{subsec:Properties-of-the}

Continuing to focus on the cold plasma, we consider the changes to
both the MCP and ECP matrices upon displacing the particle lattice
by $\epsilon$ relative to the grid. Taking $\xi_{\alpha0}\to\xi_{\alpha0}+\epsilon$,
obtaining for MCP,
\begin{equation}
M_{\alpha\beta}(\epsilon)=-\Delta^{2}\sum_{ij}K(x_{i}-\xi_{\alpha0}-\epsilon)G_{0,i-j}K'(x_{j}-\xi_{\beta0}-\epsilon)\label{eq:M-matrix-with-eps-MCP-2}
\end{equation}
for the MCP case and
\begin{equation}
M_{\alpha\beta}(\epsilon)=-\Delta^{2}\sum_{ij}K'(x_{i}-\xi_{\alpha0}-\epsilon)L_{i-j}K'(x_{j}-\xi_{\beta0}-\epsilon)\label{eq:M-matrix-with-eps-ECP-2}
\end{equation}
for the ECP case. (The special case of Sec.~\ref{sec:Block-circulant}
had $\epsilon=N_{ppc}q=\Delta$.)

In the integral form for the MCP case, Eq.~(\ref{eq:M^=00007Bmc=00007D_Integral}),
the substitutions $x'=x-\epsilon,\,y'=y-\epsilon$ show that the matrix
elements $M_{\alpha\beta}$ are independent of $\epsilon$, implying
that any variation in the matrix elements of either matrix for $N_{g}<\infty$
as $\epsilon$ varies is due to the presence of a grid, which leads
to trapezoidal rule error, i.e.~aliasing. The results of Sec.~\ref{sec:Block-circulant}
imply, for $N_{g}<\infty$, 
\begin{equation}
M_{\alpha\beta}(\Delta)=M_{\alpha\beta}(0)\label{eq:Matrix-periodDelta}
\end{equation}
for both the MCP and ECP cases. That is, the matrix elements of $M(\epsilon)$
are periodic in $\epsilon$ of period $\Delta$. Numerical results
show that, indeed, the matrix elements are periodic in $\epsilon$
of period $\Delta$, but the eigenvalues of $M(\epsilon)$ (MCP and
ECP) are periodic with a smaller period $q=\Delta/N_{ppc}<\Delta$. 

To understand this periodicity of eigenvalues of period $q$ in $\epsilon$,
let us use Eq.~(\ref{eq:M-matrix-with-eps-ECP-2}) (for MCP) to conclude
\[
M_{\alpha\beta}(q)=-\Delta^{2}\sum_{ij}K(x_{i}-\xi_{\alpha+1,0})G_{0,i-j}K'(x_{j}-\xi_{\beta+1,0})
\]
or
\begin{equation}
M_{\alpha\beta}(q)=M_{\alpha+1,\beta+1}(0),\label{eq:eps=00003DqSymmetry}
\end{equation}
and the same conclusion holds for the ECP case. Next, we write the
eigenvalue equation for eigenvectors $\delta\boldsymbol{\xi}$ and
eigenvalues $\lambda$
\begin{equation}
M\delta\boldsymbol{\xi}=\lambda\delta\boldsymbol{\xi}\label{eq:EigenvalueEq-shift}
\end{equation}
 The periodic index shift operator $Q$ is defined as $(Q\delta\boldsymbol{\xi})_{\alpha}\doteq\delta\boldsymbol{\xi}_{\alpha+1}$
. The matrix $Q$ is a permutation matrix, so it is unitary, $Q^{T}=Q^{-1}$.
Writing $\delta\boldsymbol{\xi}=Q^{T}\delta\boldsymbol{\eta}$, the
eigenvalue equation is written as 
\begin{equation}
M'\delta\boldsymbol{\eta}=\lambda\delta\boldsymbol{\eta},\label{eq:Perm-symmetry-1}
\end{equation}
\begin{equation}
M'\doteq QMQ^{T}=QMQ^{-1}\label{eq:Permut-sym-2}
\end{equation}
showing that $M'=M(q)$ has the same eigenvalues as $M(0)$.

In summary, we have shown that the matrix elements $M_{\alpha\beta}(\epsilon)$
are periodic of period $\Delta=N_{ppc}q$ in $\epsilon$ and the eigenvalues
of $M(\epsilon)$ are periodic of period $q=\Delta/N_{ppc}$ in $\epsilon$,
for both the MCP and ECP discretizations. Recall from Sec.~\ref{subsec:Dimensional-reduction-of}
that the dimensionally reduced matrix is periodic in $\epsilon$ of
period $q$, and we therefore expect the same of the eigenvalues.
We present numerical results related to these issues in Sec.~\ref{subsec:Stability-results-for}.

\subsection{Eigenvalue properties for MCP and ECP}
\label{subsec:Eigenvalue-properties-for}

We showed in Eq.~(\ref{eq:DFT-eigenvalues}) an expression for the
eigenvalues of the circulant $N_{g}\times N_{g}$ matrix $N$, with
$z=(z_{m})_{n}=e^{2\pi imn/N_{g}}$. This consists of $N_{g}$ equally
spaced values on the period for each $m$, with $-N_{g}/2+1\leq m\leq N_{g}/2$,
assuming that $N_{g}$ is even. In this case there is a Nyquist mode
with $z=-1$ as well as a neutral mode, with eigenvalue zero, for
$z=1$ ($m=0$); the latter mode, which has zero eigenvalue for both
MCP and ECP, corresponds to the zero eigenvalue mode of $M$ of Sec.~\ref{sec:Discrete-equations-of},
$\delta\boldsymbol{\xi}=\boldsymbol{c}_{p}$. The Nyquist mode is
neutral for MCP but not for ECP. The coefficients $a_{n}$ are real;
for MCP, Eq.~(\ref{eq:DFT-eigenvalues}) suggests that eigenvalues
corresponding to $z$, on the unit circle and off the real line with
the exception of the values $z=\pm1$, are complex. The eigenvalues
corresponding to these values of $z$ occur in complex conjugate pairs:
for $z\to z^{*}$, we have $\lambda\to\lambda^{*}$. For complex eigenvalues
$\lambda=\omega^{2}$, this complex conjugate property implies one
each pair of modes is unstable.

For ECP, on the other hand, this property still holds but with real
eigenvalues, so that the eigenvalues corresponding to $z$ and $z^{*}$
are degenerate. Contrast this with the generic real symmetric matrix,
which is nondegenerate. For such a matrix, a symmetry breaking perturbation
must be sufficiently large for two distinct real eigenvalues to collide
and then go off into the complex plane. Thus, for a typical symmetric
matrix there is a threshold value of the size of a symmetry breaking
perturbation above which complex conjugate roots, and instability,
can occur. Below this threshold only real eigenvalues exist. In contrast,
because of the degeneracy of the eigenvalues in $\hat{M}$ (and therefore
in $M$), the symmetric matrix of ECP can develop complex conjugate
eigenvalues, and instability, with an arbitrarily small symmetry breaking
perturbation.\footnote{For the ECP case, the dimensional reduction $M=K'^{T}LK'\to\hat{M}=LK'K'^{T}$
has $\hat{M}^{T}=K'(LK')^{T}=K'K'^{T}L$, because $L$ is symmetric.
However, the $N_{g}\times N_{g}$ matrices $K'K'^{T}$ and $L$ are
circulant, and circulant matrices, all having the same Fourier eigenvectors,
commute, showing $\hat{M}^{T}=LK'K'^{T}=\hat{M}$, showing that the
reduced matrix $\hat{M}$ is symmetric.}

The antisymmetric part of reduced MCP matrix $\hat{M}$ scales as
$1/N_{ppc}$ (or smaller, depending on the choice of kernel) for large
$N_{ppc}$, with the growth rate of the instability $\gamma$ scaling
in the same manner in MCP, for several of the modes with $z$ on the
unit circle with $z\neq\pm1$.

\subsection{Stability results for MCP}
\label{subsec:Stability-results-for}

\begin{figure}
  \centering
  \includegraphics[width=0.45\textwidth]{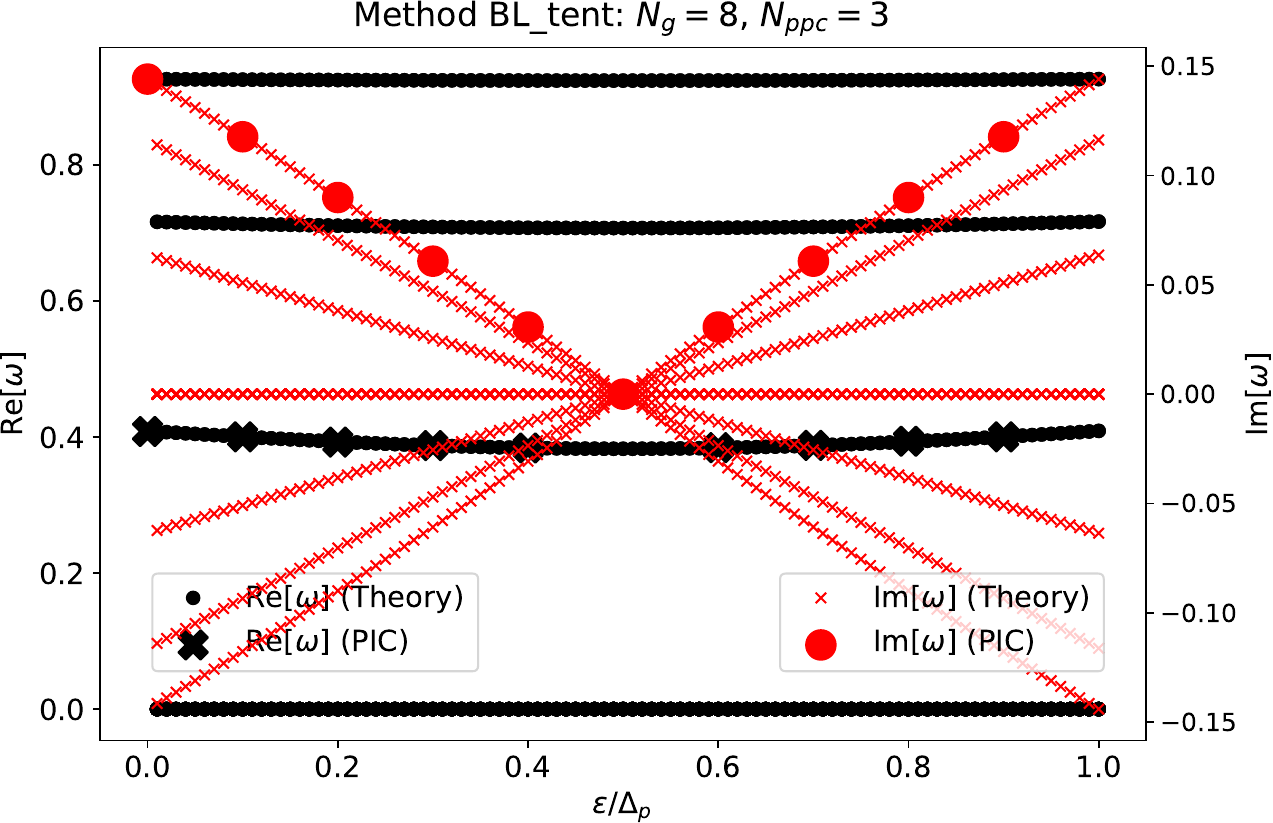} \hspace{5pt}
  \includegraphics[width=0.45\textwidth]{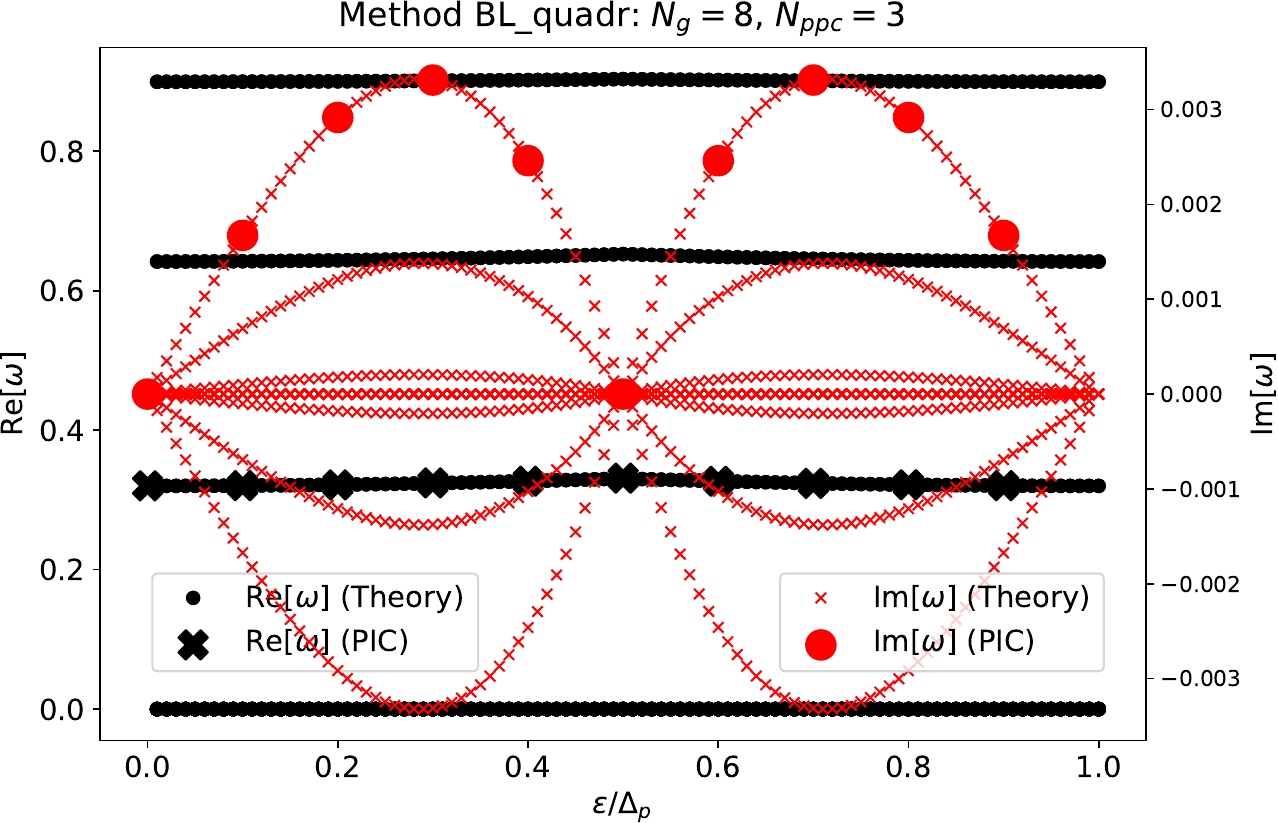}
\caption{Real and imaginary parts of $\omega$, where $\lambda=\omega^{2}$
are the eigenvalues of $q\hat{M}$, as functions of the lattice displacement
$\epsilon$ relative to $q$. Results are with $N_{g}=8,\,\,N_{ppc}=3$
for (a) linear kernel and (b) quadratic spline kernel. Values for
$\omega_{r}$ and $\gamma=\omega_{i}$ found by a PIC code are shown
as dots for the fastest growing mode.}
\label{Real-and-imaginary}
\end{figure}

In Fig.~\ref{Real-and-imaginary} we show the real and imaginary
parts of $\omega$, where eigenvalues of $\hat{M}$ have $\lambda=\omega^{2}$,
as functions of $\epsilon$. These are plotted over the period $0\leq\epsilon\leq q$,
with $N_{g}=8$ and $N_{ppc}=3$. The results are for the linear quadratic
spline kernels. The eigenvalues of $\hat{M}$ are computed by DFT
methods, and show the predicted period $q$ in $\epsilon$. For the
parameters of Fig.~\ref{Real-and-imaginary}, and for both kernels,
there are three unstable modes for $0\leq\epsilon\leq q/2$, with
$z_{m}=e^{2\pi imx_{n}}=(e^{\pi i/4},\,i,e^{3\pi i/4})$, with growth
rates $\gamma=\omega_{i}$ increasing in this order. The real frequencies
$\omega_{r}$ are observed to decrease in this order, consistent with
the decrease in frequency as wavenumber increases, according to Eq.~(\ref{eq:Cold-DispReln}).
(For these kernels, $\tilde{K}_{2}(k)$ decreases for small $k$.)
For each of the unstable modes, there is a complex conjugate damped
mode with the same real frequency, as noted in Sec.~\ref{subsec:Eigenvalue-properties-for}.
At $\epsilon=q/2,$ the stable and unstable modes cross, again leading
to three unstable modes for $q/2<\epsilon\leq q$ . Note that, for
the quadratic kernel, the growth rates are continuous with $\epsilon$
and satisfy $\gamma=0$ at $\epsilon=0$ and $\epsilon=q/2$. For
the linear kernel, we have $\gamma=0$ at $\epsilon=q/2$ but there
is a step discontinuity, with $\gamma=0$, at $\epsilon=0$.

The eigenvalues $\lambda=\omega^{2}$ shown in Fig.~\ref{Real-and-imaginary}
indeed show periodicity in $\epsilon$ of period $q$, and the matrix
elements of the dimensionally reduced matrix show the same period
(not shown.)

The results in Fig.~\ref{Real-and-imaginary} show that the Nyquist
mode for MCP is a neutral mode, with $\lambda=\omega^{2}=0$. The uniform
mode $\delta\boldsymbol{x}\propto\boldsymbol{c}_{g}$ is also a neutral
mode. Both of these observations are consistent with the analysis
of Sec.~\ref{subsec:Eigenvalue-properties-for}.

Superimposed on these values of $\omega_{r}$ and $\gamma=\omega_{i}$
are values found by a standard momentum-conserving PIC code\cite{BLBook}. Figure \ref{Real-and-imaginary}
shows the fastest growing mode, although other modes were also found
to be in agreement. The agreement with the analytic theory for both
kernels is excellent.

\begin{figure}
  \centering
  \includegraphics[width=0.45\textwidth]{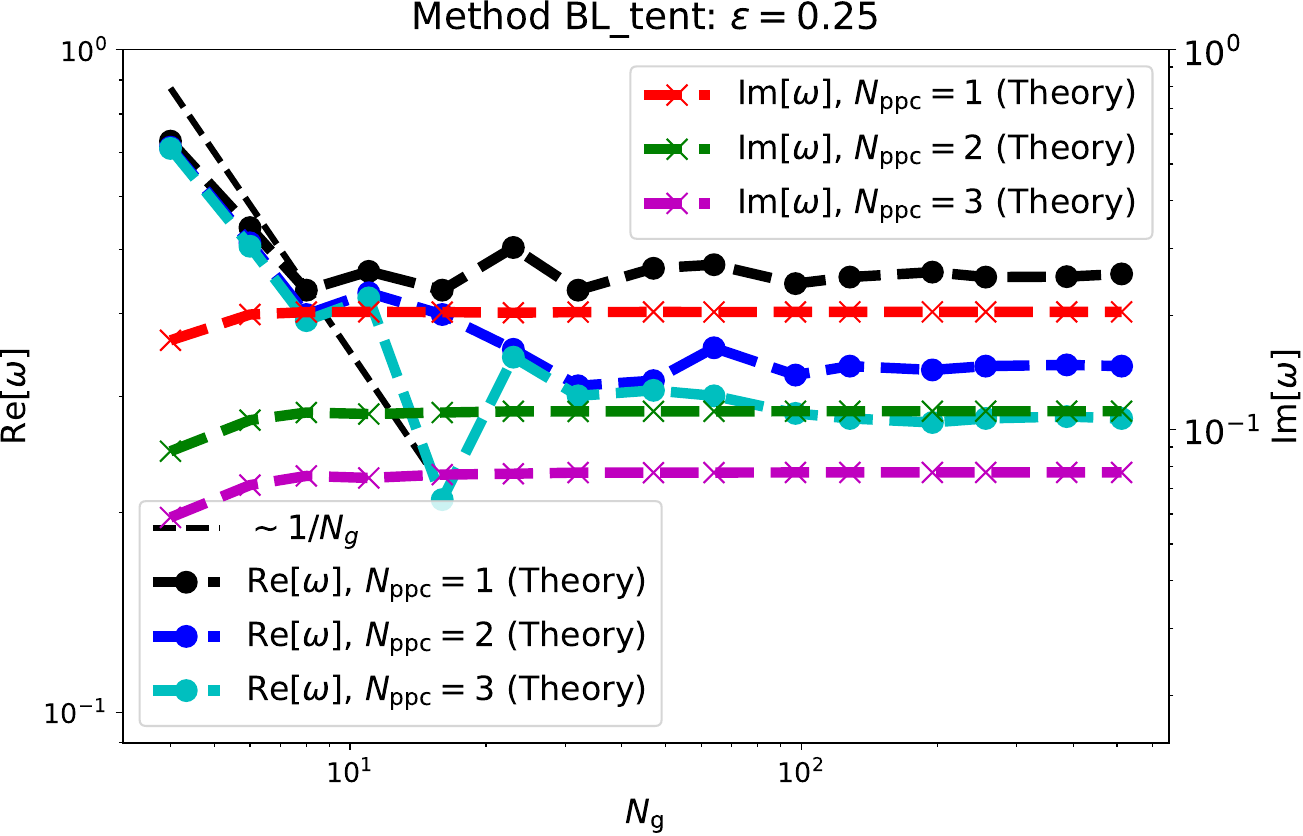} \hspace{5pt}
  \includegraphics[width=0.45\textwidth]{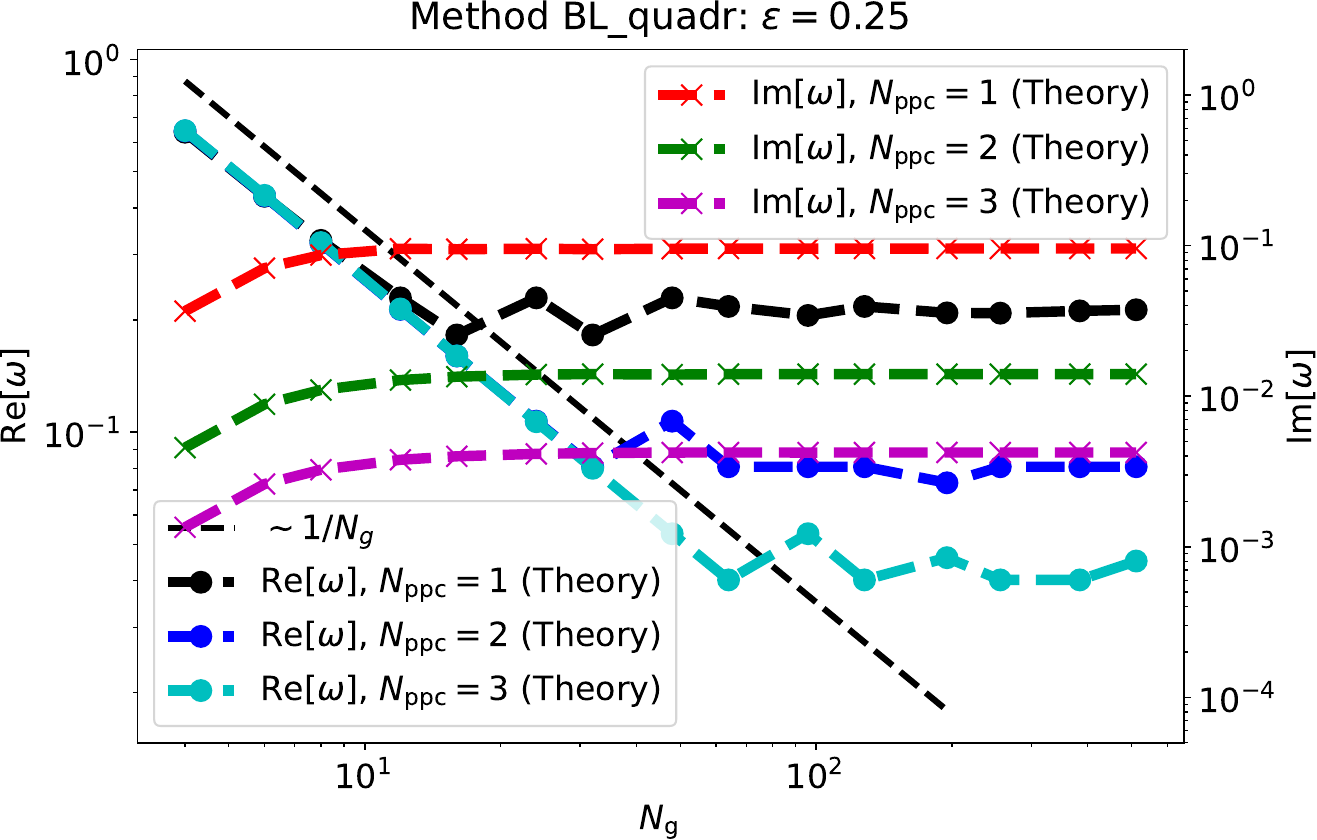}
\caption{Real and imaginary parts of $\omega$ vs.~$N_{g}$ for $N_{ppc}=1,\,2,\,3$
with (a) a linear (tent) spline and (b) a quadratic spline. This shows
convergence to a positive value of $\gamma=\omega_{i}$, and similarly
for $\omega_{r}$. This convergence to nonzero values is because of
the property $w\propto\Delta$.}
\label{Real-and-imaginary-1}
\end{figure}

Figure \ref{Real-and-imaginary-1} shows $\omega_{r}$ and $\gamma=\omega_{i}$
as functions of $N_{g}$ with fixed $N_{ppc}$, showing convergence
to a positive constant for large $N_{g}$. There are oscillations
for $N_{g}\lesssim100$. 

\begin{figure}
  \centering
  \includegraphics[width=0.42\textwidth]{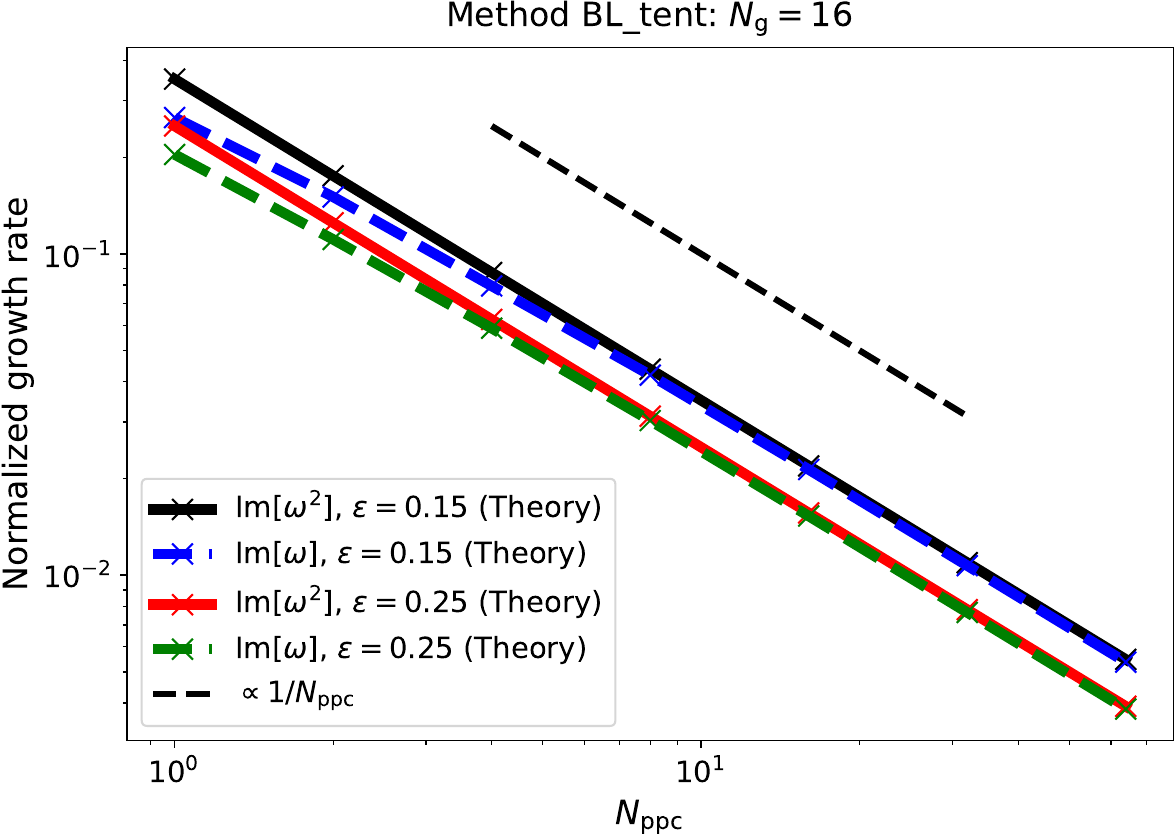} \hspace{20pt}
  \includegraphics[width=0.42\textwidth]{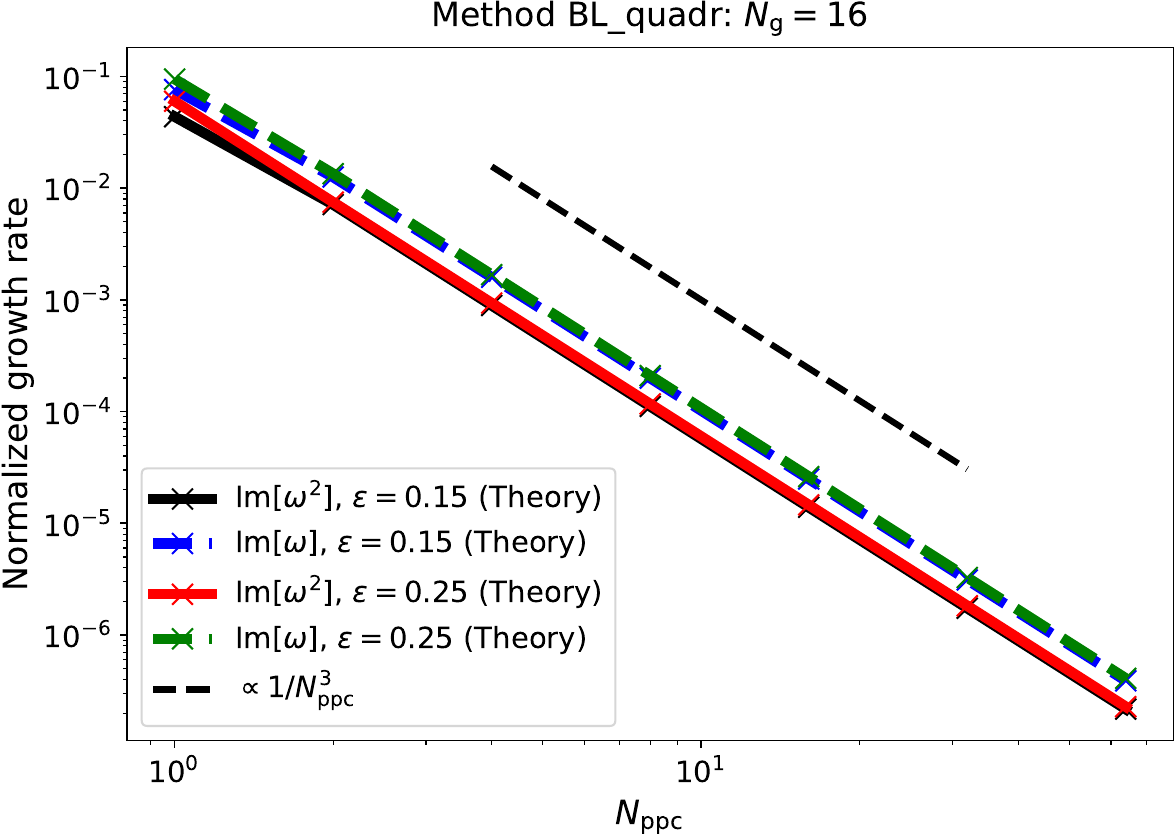} \hspace{10pt}
\caption{Growth rate $\gamma=\omega_{i}$ vs.~$N_{ppc}$ for (a) linear (tent)
spline, $N_{g}=16$, and two values of $\epsilon$ relative to $q$.
Results show $\gamma\propto1/N_{ppc}$. Same quantities in (b) for
the quadratic spline, showing $\gamma\propto1/N_{ppc}^{3}$. Cubic
spline results are similar to the quadratic, with the same scaling.}
\label{fig.:gamma-vs-Nppc}
\end{figure}

In Fig.~\ref{fig.:gamma-vs-Nppc}(a) we show the growth rate $\gamma$
(normalized as\textcolor{red}{{} }in Sec.~\ref{sec:Meshfree-formulation})
as a function of $N_{ppc}$ for two values of $\epsilon/q$, $0.15$
and $0.25$, respectively. Results are for the linear (tent) spline
kernel and $N_{g}=16$. Results using the quadratic spline are shown
in Fig.~\ref{fig.:gamma-vs-Nppc}(b), consistent with $\gamma\propto1/N_{ppc}^{3}$.
Results (not shown) with a cubic spline kernel also show $\gamma\propto1/N_{ppc}^{3}$. 

\section{Cold beam $v_{0}\protect\neq0$}
\label{sec:Cold-beam}

For a cold beam, the steps as in Sec.~\ref{subsec:Discrete-equations-for-momentum}
lead, with unperturbed orbits $\xi_{\alpha0}(t)=\xi_{\alpha0}(0)+v_{0}t$
and the MCP discretization, to 
\begin{equation}
M_{\alpha\beta}(t)=-\Delta^{2}\sum_{ij}K(x_{i}-\xi_{\alpha0}(0)-v_{0}t)G_{0,i-j}K'(x_{j}-\xi_{\beta0}(0)-v_{0}t).\label{eq:M-matrix-with-eps-MCP-1}
\end{equation}
 For the ECP case of Sec.~\ref{subsec:Discrete-equations-for} we
find, similarly,
\begin{equation}
M_{\alpha\beta}(t)=-\Delta^{2}\sum_{ij}K'(x_{i}-\xi_{\alpha0}(0)-v_{0}t)L_{i-j}K'(x_{j}-\xi_{\beta0}(0)-v_{0}t),\label{eq:M-matrix-with-eps-ECP-1}
\end{equation}
identical to the matrices in Eqs.~(\ref{eq:M-matrix-with-eps-MCP-2}),
(\ref{eq:M-matrix-with-eps-ECP-2}), but with $\epsilon\to v_{0}t$.
In Sec.~\ref{sec:Shifted-lattice}, we showed that, for the meshfree
case, the variation with respect to $\epsilon$ disappears. It is
also true that in the integral forms of Eqs.~(\ref{eq:M-matrix-with-eps-MCP-1}),
(\ref{eq:M-matrix-with-eps-ECP-1}), Eqs.~(\ref{eq:M^=00007Bmc=00007D_Integral})
and (\ref{eq:Integral-form-MCP-1}) with $\xi_{\alpha0}\to\xi_{\alpha0}+v_{0}t$,
there is no variation with respect to time. For $N_{g}<\infty$, on
the other hand, oscillation of $M_{\alpha\beta}(\epsilon)$ for $v_{0}=0$
with period $q$ in $\epsilon$ is observed; these results are discussed
in Sec.~\ref{subsec:Properties-of-the} and shown in Sec.~\ref{subsec:Stability-results-for}.
This variation is due to the trapezoidal rule error, or equivalently
to aliasing. The results showing periodicity of the dynamical matrix
elements of period $\Delta$ in $\epsilon$ imply periodicity of the
matrices in Eqs.~(\ref{eq:M-matrix-with-eps-MCP-1}), (\ref{eq:M-matrix-with-eps-ECP-1})
in time. The period of this oscillation $T=\Delta/v_{0}$ is the time
for a particle of velocity $v_{0}$ to pass one grid cell.  That is,
the fundamental frequency is $\Omega=2\pi v_{0}/\Delta=k_{g}v_{0}$,
where $k_{g}=2k_{N}$ is the grid wavenumber and $k_{N}$ is the Nyquist
wavenumber, with period $2\Delta$. 

This oscillation in time of the matrix elements of $M$ at frequency
$\Omega=k_{g}v_{0}$ for $v_{0}\neq0$ leads to the possibility of
parametric instabilities, for both the MCP and ECP forms. The fundamental
Mathieu resonance is where the driving frequency $\Omega$ is twice
the natural frequency given by the dispersion relation.

For a warm plasma, for each wavenumber $k$ there is a class of particles
with velocity $v_{0}$ such that the natural frequency is in resonance
with the oscillation. We will discuss these issues in depth in a forthcoming
paper. Preliminary results suggest that such instabilities can occur
due to a resonant class of particles, and the nonresonant particles
can be dealt with in a continuum manner, giving the Langmuir dispersion
relation, modified by the kernel but, as noted, with zero temperature, in Eq.~(\ref{eq:Cold-DispReln}).
According to this suggestion, instability is described by a crossing
of the Langmuir dispersion relation with the Mathieu-like resonant
frequency. This crossing should disappear, stabilizing the mode, with
a large enough temperature (large enough Debye length.) If there is
wavenumber crossing, but it is outside the Nyquist range, it will
be aliased into that range, a second occurrence of aliasing for such
grid instabilities. We will deal with these issues in detail in a
forthcoming publication.

\section{Summary and Discussion}
\label{sec:Summary-and-Discussion}

We have analyzed standard PIC methods for instabilities related to
the presence of a grid ($N_{g}<\infty$) and to a finite number of
macroparticles ($N_{p}<\infty$). To do so, we have started with a
meshfree formulation for a $1D$ electrostatic plasma periodic in
$x$ with period $1$, with electrons represented by the $N_{p}$
macroparticles and having a immobile neutralizing background of ions.
In this analysis, each macroparticle was characterized by a specific
kernel or particle shape of a fixed width, and the density is computed
by means of kernel density estimation\cite{BLBook,HockneyEastwood,Bishop}.
The force on a macroparticle was written as the integral of the electric
field over the macroparticle charge density, i.e.~the particle (or
kernel) shape. We have discussed briefly the invariants associated
with this formulation, and the exact equivalence of the Vlasov-Gauss
formulation in terms of the electric field, the Vlasov-Poisson formulation
in terms of a scalar potential $\phi$, and the Vlasov-Amp\`ere approach.

The next step discussed was linearization of these meshfree equations,
first for a cold plasma. We started by introducing a macroparticle
lattice $\{\xi_{\alpha0}\}_{\alpha=0}^{N_{p}-1}$ with $u_{\alpha0}=0$,
describing an exact equilibrium, with uniform density, zero electric
field, no flow and zero temperature. We then derived linearized meshfree
equations of motion for particles, in terms of a dynamical matrix
for the perturbed particle orbits about the lattice, with the field
solve taken into account. This dynamical matrix is symmetric positive
definite (SPD) because the same kernel $K(x)$ was used to compute
the density and electric field, and to compute the force on a macroparticle.
We showed the dispersion relation for this system, the Langmuir dispersion
relation, modified by the kernel shape and width $w$.

The next step involved discretizing to a grid $\{x_{i}\}_{i=0}^{N_{g}-1}$
for $N_{p}/N_{g}=N_{ppc}$, a positive integer. We derived the dynamical
matrix for linearized particle motions in the presence of a self-consistent
electric field given on the grid. We considered two discretizations,
the first of which is in terms of the electric field and is of them
used in momentum conserving PIC codes, MCP. The other is in terms
of the electrostatic potential, and appears in energy conserving PIC
formulations, ECP. We showed that this matrix is symmetric positive
definite (SPD) for ECP, again because of the use of the same kernel
for density estimation and for computing the force on a macroparticle.
The dynamical matrix for MCP was found not to have this SPD property. 

We have discussed the block-circulant property of the $N_{p}\times N_{p}$
dynamical matrix $M$, and a process we called dimensional reduction,
to a $N_{g}\times N_{g}$ matrix $\hat{M}$. This latter matrix is
circulant and can be analyzed by DFT methods. The circulant property
means that this reduced matrix is normal, and we discussed the possibility
that the original dynamical matrix is not normal with respect to the
$(l^{2})_{N_{p}}$ inner product.

These results showed, for a cold plasma in MCP discretization, that
the dynamical matrix varies with displacement of the lattice relative
to the grid $\xi_{\alpha0}\to\xi_{\alpha0}+\epsilon$, leading to
variation of the eigenvalues of $M$ with the displacement $\epsilon$.
For MCP this variation is due to numerical errors in the trapezoidal
rule integration, or equivalently aliasing. This variation can yield
eigenvalues in complex conjugate pairs, leading instability. For the
ECP discretization, the matrix elements vary with $\epsilon$ in a
similar manner as for the MCP discretization. However, the eigenvalues
are real and non-negative, due to the SPD nature of the ECP matrix.

We have determined the scaling of the dynamical matrix elements and
eigenvalues with $N_{g}$, the number of grid points and with $N_{ppc}$,
the number of particles per cell. We have discussed aliasing and its
equivalence to trapezoidal rule integration error in Sec.~\ref{sec:Properties-of-the}
and in Appendix \ref{app:Trapezoidal}. In Sec.~\ref{sec:Properties-of-the}
and Appendix \ref{app:DimReduction} we discuss the above method of
dimensional reduction to be applied to the dynamical matrix, for both
the ECP and MCP discretizations. This reduction leads to circulant
matrices, which can be easily analyzed by discrete Fourier transform
(DFT) methods.

We have observed the close relationship between the behavior of the
matrices $M_{\alpha\beta}$ (MCP and ECP) for a cold plasma with $\epsilon$
and the variation with time for a cold drifting beam. Results related
to instabilities in these two systems will appear in a subsequent
publication, as well as the subject of grid instabilities for a warm
plasma, based on the cold beam (MCP and ECP) results.

\section*{Acknowledgments}

We wish to thank B.~Shadwick for useful discussions. The work of
EGE was supported by Sandia National Laboratories 2026 New Mexico
Small Business Assistance Program, Project \#16725.
Sandia National Lab-
oratories is a multi-mission laboratory managed and
operated by National Technology \&
Engineering Solutions of Sandia, LLC (NTESS), a wholly owned subsidiary of
Honeywell International Inc., for the U.S. Department of Energy's
National Nuclear Security Administration (DOE/NNSA) under contract DE-NA0003525.
This written work is authored by an employee of NTESS. The employee, not NTESS,
owns the right, title and interest in and to the written work and is
responsible for its contents. Any subjective views or opinions that
might be expressed in the written work do not necessarily represent
the views of the U.S. Government. The publisher acknowledges that the
U.S. Government retains a non-exclusive, paid-up, irrevocable,
world-wide license to publish or reproduce the published form of this
written work or allow others to do so, for U.S. Government
purposes. The DOE will provide public access to results of federally
sponsored research in accordance with the DOE Public Access Plan.

\appendix

\section{Trapezoidal error and aliasing}
\label{app:Trapezoidal}

The linearized MCP and ECP discretizations lead to the $N_{p}\times N_{p}$
matrices $M$ of Eqs.~(\ref{eq:Dyn-Mx-BL-1}) and (\ref{eq:Mab-ECP}),
respectively. These are the trapezoidal rule approximations to the
integrals such as that in Eq.~(\ref{eq:M^=00007Bmc=00007D_Integral}).
To analyze the errors in these discretizations, we start with an analytic
periodic function $f$ on $[0,1]$. We will relax the analyticity
condition later, allowing steps in $f$ or its derivatives at $x=0$
($x=1$). The error in its trapezoidal rule formula for integration,
the one dimensional version of the discretization used in Sec.~\ref{sec:Discrete-equations-of},
starts with the Taylor series $f(x)=f(x_{i})+(x-x_{i})f'(x_{i})+(x-x_{i})^{2}f''(x_{i})/2+\cdots$
for $x_{i}\leq x<x_{i+1}$ and leads to
\[
\int_{0}^{1}f(x)dx=\sum_{i=0}^{N-1}\int_{x_{i}}^{x_{i+1}}f(x)dx
\]
\begin{equation}
=\Delta\sum_{i=0}^{N-1}f(x_{i})-E(\Delta),\label{eq:TrapWithError}
\end{equation}
where, for application to Sec.~\ref{sec:Discrete-equations-of},
we have $N=N_{g}$. This is a form of backward error analysis, and
$E(\Delta)$ is the error in approximating the sum as an integral,
the negative of approximating the integral as a sum. This formula, for
application to periodic functions $f$, has $f(x_{N})=f(x_{0})$,
so that the $i=0$ term can be written as $\Delta f(x_{0})=(\Delta/2)\left(f(x_{0})+f(x_{N})\right)$,
giving the standard trapezoidal expression, with the remainder or
error term $-E(\Delta)$. We have, to order $\Delta^{3}$ (with $N_{g}=1/\Delta$),
\begin{equation}
E(\Delta)=-\frac{\Delta^{2}}{2}\sum_{i}f'(x_{i})-\frac{\Delta^{3}}{6}\sum_{i}f''(x_{i}).\label{eq:TrapRemainder}
\end{equation}
 We also have
\begin{equation}
\int_{0}^{1}f'(x)dx=\Delta\sum_{i}f'(x_{i})+\frac{\Delta^{2}}{2}\sum_{i}f''(x_{i})\label{eq:Trap-for-f'}
\end{equation}
with 
\begin{equation}
\int_{0}^{1}f''(x)dx=\Delta\sum_{i}f''(x_{i}).\label{eq:Aps-integral-of-f''}
\end{equation}
Using Eqs.~(\ref{eq:TrapRemainder}-\ref{eq:Aps-integral-of-f''})
we find
\begin{equation}
E(\Delta)=-\frac{\Delta}{2}\int_{0}^{1}f'(x)dx+\frac{\Delta^{2}}{12}\int_{0}^{1}f''(x)dx.\label{eq:R(Delta)-to-O(Delta^2)}
\end{equation}

At this point we relax the analyticity assumption. If $f$ has a jump
at a single value $x=x_{0}$ in $[0,1]$, we can shift $x\to x-x_{0}$
to have the jump occur at $x=0$ ($x=1$) and we find, to lowest order,
\[
E(\Delta)=-\frac{\Delta}{2}\left(f(1-)-f(0+)\right)=\frac{\Delta}{2}[f]_{0},
\]
where $[\cdot]_{0}$ is the jump at $x=0$. If $f$ is continuous
but $f'$ has a jump at $x=0$ the lowest order error is 
\[
E(\Delta)=\frac{\Delta^{2}}{12}\left(f'(1-)-f'(0+)\right)=-\frac{\Delta^{2}}{12}[f']_{0}.
\]
The generalization to functions with derivatives up to $f^{[k-1]}$
continuous but with a jump in $f^{[k]}$ is 
\begin{equation}
E(\Delta)\propto\Delta^{k+1}[f^{[k]}]_{0}.\label{eq:TrapErrorResultScaling}
\end{equation}

In Fourier space, the above results are related to the well-known
results regarding Fourier series. These are: If $f$ has a step, its
Fourier series has $\tilde{f}_{m}\propto1/m$; if $f$, ... $f^{[k-1]}$
are continuous but $f^{[k]}$ has a step, then we have $\tilde{f}_{m}\propto1/m^{k+1}$. 

So far, we have taken the special case for which $f$ is analytic
on $(0,1)$ with jumps in $f^{[k]}$ at the endpoints, but these results
are valid for jumps at a number of points in $[0,1]$.

For applying to data on a grid (samples) with periodic boundary conditions,
we start with the Fourier series for $f$,
\begin{equation}
f(x)=\sum_{m=-\infty}^{\infty}a_{m}e^{2\pi inx},\,\,\,a_{n}=\int_{0}^{1}f(x)e^{-2\pi inx}dx.\label{eq:FSeries}
\end{equation}
We then replace this by a Discrete Fourier Transform (DFT) at samples
$x_{j}$ and discrete wavenumbers $k_{m}=2\pi m$,
\begin{equation}
f(x_{j})=\sum_{m\in\mathcal{N}}b_{m}e^{2\pi imx_{j}},\,\,\,b_{m}=\frac{1}{N}\sum_{j=0}^{N-1}f(x_{j})e^{-2\pi imx_{j}}.\label{eq:DFT}
\end{equation}
Here, $\mathcal{N}$ stands for the $N$ integers in the Nyquist range
$m=-m_{N}+1,\cdots,m_{N}$ (for $N$ even), where $m_{N}=N/2$ is
the Nyquist mode number. We also have $x_{j}=j\Delta=j/N$ for $j=0,\dots,N-1$.
Evaluating $f(x_{j})$ from Eq.~(\ref{eq:FSeries}) and substituting
in $b_{m}$ from Eq.~(\ref{eq:DFT}) we find, using the identity
\[
\sum_{j=1}^{N_{g}}e^{2\pi(n-m)ix_{j}}=N\sum_{p=-\infty}^{\infty}\delta_{n-m-Np},
\]
the \emph{aliasing formula}
\begin{equation}
b_{m}=\sum_{p=-\infty}^{\infty}a_{m+Np}.\label{eq:FS->DFT-WithAlias}
\end{equation}
We have $b_{m}=a_{m}+A_{m}$, where 
\begin{equation}
A_{m}=\sum_{p\neq0}a_{m+Np}\label{eq:DFTAlias}
\end{equation}
is the alias, the error due to the function $f$ having nonzero amplitudes
$a_{n}$ for wavenumbers outside the Nyquist range $\mathcal{N}$.
For power-law behavior $|a_{m}|\sim1/m^{k}$, we can find the scaling
for the alias or trapezoidal rule error, specialized to $m=0$:

\begin{equation}
E(\Delta)=A_{0}=\sum_{p\neq0}\frac{1}{(Np)^{k}}\propto\Delta^{k}.\label{eq:DFTAlias-final-scaling}
\end{equation}
These results hold only if the sum converges. The sum converges unconditionally
for $k>1$. For $k=1$, i.e.~when $f$ has a jump discontinuity,
the terms typically oscillate, and the sum converges conditionally.
(This, of course, means that the value of the sum depends on the order
of the terms, but it is known\cite{stein2003fourier} that the symmetric
sum order $p=0,\,1,\,-1,\,2,\,-2,\dots$ provides the most accurate
result.)

Comparing this aliasing error with the trapezoidal error in Eq.~(\ref{eq:TrapErrorResultScaling}),
we note that for $f^{[k-1]}$ continuous but having a jump in $f^{[k]}$,
the trapezoidal error scales as $E(\Delta)\propto\Delta^{k+1}$. For
the same function $f$, the Fourier coefficients scale as $a_{m}\sim1/m^{k+1}$,
giving an aliasing error in the $m=0$ sum $E(\Delta)\propto\Delta^{k+1}$,
in agreement with the trapezoidal error results.

Consider the double integrals that define the matrix elements $M_{\alpha\beta}$,
in either the MCP or the ECP forms of Eqs.~(\ref{eq:M^=00007Bmc=00007D_Integral})
and (\ref{eq:Integral-form-MCP-1}) are of the form
\begin{equation}
M=\Delta^{2}\sum_{ij}f(x_{i})N(x_{i}-x_{j})g(x_{j}),\label{eq:Triple-matrix-general-form}
\end{equation}
where $f(x)=K(x-\xi_{\alpha}),\,N(x)=G_{0}(x),\,g(x)=K'(x-\xi_{\beta})$
for MCP, and $f(x)=K'(x-\xi_{\alpha}),\,N(x)=L(x),\,g(x)=K'(x-\xi_{\alpha})$
for ECP. For the purpose of finding the Fourier series to evaluate
the aliasing in Eq.~(\ref{eq:Triple-matrix-general-form}), the continuous
form for the 2D $m=0$ Fourier series coefficient is
\begin{equation}
M=\int dxdyf(x)N(x-y)g(y).\label{eq:Triple-continuum}
\end{equation}
Let us suppose $[f^{(k_{1})}]\neq0$, i.e.~$f_{m}\sim1/m^{k_{1}+1}$
and similarly, $N_{m}\sim1/m^{k_{2}+1}$ and $g_{m}\sim1/m^{k_{3}+1}$.
Also note that the quantity $M$ in Eq.~(\ref{eq:Triple-continuum})
is symmetric with respect to permuting $f,\,L,$ and $g$. Let us
assume that $f$ is the least smooth\footnote{The functions can be permuted in ways other than the least smooth
in the first position, and the results of a much more intricate derivation
with these other orderings show the same results.} of these three functions, i.e.~$k_{1}\leq k_{2},\,k_{3}$. Then,
$M$ can be put in the form
\[
\int f(x)h(x)dx,
\]
where $h(x)=\int dyN(x-y)g(y)$ is a convolution. The convolution
has Fourier coefficients
\[
h_{m}=N_{m}g_{m}\sim1/m^{k_{2}+k_{3}+2},
\]
i.e.~$h(x)$ has a jump in $[h^{(k_{2}+k_{3}+1)}]$, whereas $f$
has a jump in its $k_{1}$'th derivative. So the smoothness of $f(x)h(x)$
is dominated by $f$, meaning that the error in $M$ is of order 
\begin{equation}
E(\Delta)\sim\Delta^{k_{1}+1}.\label{eq:E-alias-powerlaw}
\end{equation}
In particular, if there is a step in $f$, leading to $k_{1}=0$,
then we have $E_{alias}\sim\Delta$.

For the linear kernel, we have the following table of values of $k_{1},\,k_{2},\,k_{3}$:
\noindent \begin{center}
\begin{tabular}{|c|c|c|}
\hline 
$\bcancel{}$ & MCP & ECP\tabularnewline
\hline 
\hline 
$k_{1}$ & 1 & 0\tabularnewline
\hline 
$k_{2}$ & 0 & 1\tabularnewline
\hline 
$k_{3}$ & 0 & 0\tabularnewline
\hline 
\end{tabular}
\par\end{center}

\noindent \begin{flushleft}
In both MCP and ECP the minimum $\min(k_{1},k_{2},k_{3})$ is zero,
implying
\par\end{flushleft}

\begin{equation}
E_{alias}\sim\Delta.\label{eq:Ealias=00003DDelta}
\end{equation}

For the quadratic spline the table is
\begin{center}
\begin{tabular}{|c|c|c|}
\hline 
$\bcancel{}$ & MCP & ECP\tabularnewline
\hline 
\hline 
$k_{1}$ & 2 & 1\tabularnewline
\hline 
$k_{2}$ & 0 & 1\tabularnewline
\hline 
$k_{3}$ & 1 & 1\tabularnewline
\hline 
\end{tabular}
\par\end{center}

For the cubic spline, this table becomes
\begin{center}
\begin{tabular}{|c|c|c|}
\hline 
$\bcancel{}$ & MCP & ECP\tabularnewline
\hline 
\hline 
$k_{1}$ & 3 & 2\tabularnewline
\hline 
$k_{2}$ & 0 & 1\tabularnewline
\hline 
$k_{3}$ & 2 & 2\tabularnewline
\hline 
\end{tabular}
\par\end{center}

For MCP we again have $\min(k_{1},k_{2},k_{3})=0$, leading again
to $E_{alias}\sim\Delta$. Note that in each table the sum $k_{1}+k_{2}+k_{3}$
is the same for MCP and ECP, because the two integrals are related
by an integration by parts. For ECP with a quadratic spline, we have
$\min(k_{1},k_{2},k_{3})=1$, yielding $E(\Delta)\sim\Delta^{2}$.
However, these estimates are less important for the ECP case because
stability is guaranteed by the SPD property. 

The error in Eq.~(\ref{eq:Ealias=00003DDelta}) is the size of the
aliasing (trapezoidal rule) error due to the discontinuities in the
functions or its derivatives. The other source of aliasing we have
discussed is the small width of the kernel $K$. If this width is
$w$, for conventional PIC we also have $w\propto\Delta$, with a
proportionality factor ($2\text{ to }4$) depending on the kernel:
Linear to cubic, the width scales as $w=2\Delta$, $w=3\Delta$, $w=4\Delta$,
where $w$ is the \emph{full} width. Furthermore, the width in $\tilde{K}$
is $m_{w}\sim1/w\sim1/\Delta\sim N$. The DFT on the $N$ grid points
has aliasing proportional to the number of aliases of $\tilde{K}(w)$.
This number of aliases (copies of the interval $[-m_{N},m_{N}]$ within
the width of $\tilde{K}(w)$, where $m_{N}=N/2$) scales as \#aliases$\sim m_{w}/m_{N}\sim N/N=1$.
That is, this aliasing error is $E(\Delta)=C_{0}=O(1)$.

Since $C_{0}$ and $C_{1}$ are the lowest order contributions, we
can add the contribution to aliasing due to the lack of kernel smoothness
and that due to the kernel width leads to the estimate
\begin{equation}
E(\Delta)\approx C_{0}+C_{1}\Delta\label{eq:E_alias-both}
\end{equation}
for the linear kernel (MCP and ECP) and for the quadratic kernel (MCP
only.) The coefficients $C_{0}$, $C_{1}$ depend on the number of
particles per cell $N_{ppc}$, as well as the particle shape. This
issue will be discussed further in a forthcoming publication.

The fact that $E(\Delta)$ does not approach zero as $\Delta\to0$
($C_{0}\neq0$) is due to the fact, noted above, that the grid spacing
$\Delta$ is comparable to the kernel width $w$, so that the kernel
(or its derivative) is not well resolved, and the limit $\Delta\to0$
does not necessarily converge to the integrals in the meshfree formulation
in Sec.~\ref{sec:Meshfree-formulation}. This point, related to $w\propto\Delta$,
is a fundamental shortcoming of the conventional PIC scheme, and an
analysis of approaches to overcome this difficulty will be discussed
in a future publication.

\section{Reduction of the lattice matrix to a grid matrix}
\label{app:DimReduction}

The dynamical matrices with elements $M_{\alpha\beta}$ are of the
form $M=K^{T}G_{0}K'$ and $M=K'^{T}LK'$, for MCP and ECP, respectively.
For $N_{p}>N_{g}$, since $K,\,\,K'$ are $N_{g}\times N_{p}$, and
$G_{0},\,\,L$ are $N_{g}\times N_{g}$, the resulting dynamical matrices
are $N_{p}\times N_{p}$ but with rank at most $N_{g}$. That is,
they have at least $N_{p}-N_{g}$ zero eigenvalues (trivial eigenvalues.)
This point makes it quite inefficient to compute eigenvalues of the
dynamical matrix coming from the MCP or ECP discretizations, especially
if $N_{p}\gg N_{g}$, i.e.~$N_{ppc}\gg1$. 

For the MCP discretization we can replace $M=K^{T}G_{0}K'$ with $(G_{0}K')\,\,K^{T}$,
using a well-known result (See Ref.~\cite{meyer2023matrix}.) This
theorem says that, for $A$ of size $N_{p}\times N_{g}$ and $B$
of size $N_{g}\times N_{p}$, with $N_{p}\geq N_{g}$, the $N_{g}$
nontrivial eigenvalues of $AB$ ($N_{p}\times N_{p}$) are equal to
the eigenvalues of $BA$ ($N_{g}\times N_{g}$), plus $AB$ has $N_{p}-N_{g}$
zero eigenvalues, the \emph{trivial eigenvalues}. (We assume all matrices
are of full rank.) Indeed, suppose we have $AB\delta\boldsymbol{\xi}=\lambda\delta\boldsymbol{\xi}$,
with $\delta\boldsymbol{\xi}\in\mathbb{R}^{N_{p}}$. Then, defining
the $\mathbb{R}^{N_{g}}$ vector $\delta\boldsymbol{x}=B\delta\boldsymbol{\xi}$,
we immediately conclude
\begin{equation}
BA\delta\boldsymbol{x}=\lambda\delta\boldsymbol{x}.\label{eq:BA-eigenvalues}
\end{equation}
Since we assumed that $A$ and $B$ are of full rank, we conclude
$\lambda\neq0$ from Eq.~(\ref{eq:BA-eigenvalues}). The trivial
eigenvalues of $AB$ correspond to the null space of $B$: Clearly,
if $B\delta\boldsymbol{\xi}=0$ then $\delta\boldsymbol{\xi}$ is
in the null space of $M=AB$. Conversely, assume $AB\delta\boldsymbol{\xi}=0$.
Defining $\delta\boldsymbol{x}=B\delta\boldsymbol{\xi}$, then we
require $A\delta\boldsymbol{x}=0$, but this is not possible for $\delta\boldsymbol{x}\neq0$
because this equation is overdetermined, since we are assuming that
$A$ is of full rank. 

For the computational results shown in Sec.~\ref{sec:Shifted-lattice},
this modification was found to be useful for avoiding the $N_{p}-N_{g}$
trivial eigenvalues, which is especially useful for $N_{p}\gg N_{g}$.
Also, writing $K_{i\alpha}=K(x_{i}-\xi_{\alpha0})$, we find 
\begin{equation}
K_{i+1,\alpha+N_{ppc}}=K_{i\alpha}.\label{eq:BC-of-bare-K}
\end{equation}
That is $K_{i\alpha}$ is block-circulant with blocks of dimensions
$1\times N_{ppc}.$ The same holds for $K'$ and $K^{T}$ (and $K'^{T}$)
are block-circulant with blocks of dimensions $N_{ppc}\times1.$ These
observations, plus the circulant nature of the matrices $G_{0,ij}$
(for the MCP case) and $L_{ij}$ (for the ECP case), lead to the block-circulant
nature of the $N_{p}\times N_{p}$ dynamical matrix $M$ (with $N_{ppc}\times N_{ppc}$
blocks.) These properties also lead to the fact that the dimensionally
reduced matrix $\hat{M}$ is circulant, allowing the simplification
of the DFT analysis of eigenvalues.

The reduction for MCP can take the form 
\begin{equation}
\hat{M}=K'\,(K^{T}G_{0})\label{eq:Reduced-D-M-MCP}
\end{equation}
also, leading to similar results. A similar reduction for ECP, $M=K'^{T}LK'\to N=LK'K'^{T}$
(or $N=K'K'^{T}L$), is possible but this is not as important because
of the SPD nature of $M$ for this case.

A related point is the connection between this dimensional reduction
and the introduction of fluid-like grid variables. Taking Eq.~(\ref{eq:KDE})
we find the perturbed density on the grid $\{x_{i}\}_{i=0}^{N_{g}-1}$
to be
\[
\delta\rho_{i}=-q\sum_{\beta}K'(x_{i}-\xi_{\beta0})\delta\xi_{\beta}
\]
or
\[
\delta\boldsymbol{\rho}=-qK'\delta\boldsymbol{\xi},
\]
so that for MCP $\delta\ddot{\boldsymbol{\xi}}=q\Delta^{2}K^{T}G_{0}K'\delta\boldsymbol{\xi}$
implies 
\begin{equation}
\delta\boldsymbol{\ddot{\rho}}=q\Delta^{2}K'K^{T}G_{0}\delta\boldsymbol{\rho},\label{eq:Grid-eq-rho}
\end{equation}
with $K'K^{T}G_{0}$, the reduced-dimension $N_{g}\times N_{g}$ matrix
of Eq.~(\ref{eq:Reduced-D-M-MCP}). This is identical to one form
of the reduced dynamical matrix $\hat{M}$ of the last paragraph,
meaning that the dimensional reduction described above as a purely
mathematical step is equivalent to the matrix of the fluid-like equations
on the $N_{g}$ grid in terms of the density on the grid.

\bibliographystyle{unsrt}
\bibliography{biblio}

\end{document}